\documentclass[graybox]{svmult}

\usepackage{mathptmx}       
\usepackage{helvet}         
\usepackage{courier}        
\usepackage{type1cm}        
\usepackage{makeidx}         
\usepackage{graphicx}        
\usepackage{multicol}        
\usepackage[bottom]{footmisc}

\makeindex             

\begin{document}

\title*{Percolation Models of Self-Organized Critical Phenomena}
\titlerunning{Percolation Models of Self-Organized Critical Phenomena}
\author{Alexander V. Milovanov}
\institute{Alexander V. Milovanov \at Associazione EURATOM-ENEA sulla Fusione, Centro Ricerche Frascati, Via Enrico Fermi 45, I-00044 Frascati (Rome), Italy, EU \email{alexander.milovanov@enea.it}
\and Also at: \at Department of Space Plasma Physics, IKI-Space Research Institute, Russian Academy of Sciences, 84/32 Profsoyuznaya street, 117997 Moscow, Russian Federation \email{amilovan@iki.rssi.ru}}
%
%
\maketitle

\abstract*{Each chapter should be preceded by an abstract (10--15 lines long) that summarizes the content. The abstract will appear \textit{online} at \url{www.SpringerLink.com} and be available with unrestricted access. This allows unregistered users to read the abstract as a teaser for the complete chapter. As a general rule the abstracts will not appear in the printed version of your book unless it is the style of your particular book or that of the series to which your book belongs.
Please use the 'starred' version of the new Springer \texttt{abstract} command for typesetting the text of the online abstracts (cf. source file of this chapter template \texttt{abstract}) and include them with the source files of your manuscript. Use the plain \texttt{abstract} command if the abstract is also to appear in the printed version of the book.}

\abstract{In this chapter of the e-book ``Self-Organized Criticality Systems" we summarize some theoretical approaches to self-organized criticality (SOC) phenomena that involve percolation as an essential key ingredient. Scaling arguments, random walk models, linear-response theory, and fractional kinetic equations of the diffusion and relaxation type are presented on an equal footing with theoretical approaches of greater sophistication, such as the formalism of discrete Anderson nonlinear Schr\"odinger equation, Hamiltonian pseudochaos, conformal maps, and fractional derivative equations of the nonlinear Schr\"odinger and Ginzburg-Landau type. Several physical consequences are described which are relevant to transport processes in complex systems. It is shown that a state of self-organized criticality may be unstable against a bursting (``fishbone") mode when certain conditions are met. Finally we discuss SOC-associated phenomena, such as: self-organized turbulence in the Earth's magnetotail (in terms of the ``Sakura" model), phase transitions in SOC systems, mixed SOC-coherent behavior, and periodic and auto-oscillatory patterns of behavior. Applications of the above pertain to phenomena of magnetospheric substorm, market crashes, and the global climate change and are also discussed in some detail. Finally we address the frontiers in the field in association with the emerging projects in fusion research and space exploration.}

\section{The percolation problem}
\label{sec:1}

The standard theory of percolation \cite{Hammer} began with an attempt to make statistical predictions about the possibility for a fluid to filter through a random medium, predictions that could be applied to a variety of physics problems, such as epidemic processes with and without immunization, the underground spread of pollution, and electrical discharges in thunderstorms. The phenomenon is characterized by a finite threshold, to be associated with a critical concentration of fractures, pores, or other sort of conducting channels in the medium, below which the spread is limited to a finite domain of ambient space, and is unlimited otherwise. The percolation problem is relevant for a number of transport problems with threshold behavior as for instance Anderson localization \cite{And} and hopping conduction in amorphous solids \cite{Efros}. The percolation transition is perhaps the simplest phase transition-like phenomenon, with the macroscopic connectedness thought as a spontaneously occurring property, and the concentration of conducting elements as the control parameter (similar to the thermodynamical temperature) \cite{Isi}.   

\subparagraph{Site and bond percolation} Given a periodic lattice, embedded in a $d$-dimensional Euclidean space, one can choose between two alternative formulations of the percolation problem: site and bond. The differences between site and bond percolation are actually very subtle and are manifest in a typically lower threshold for the bond problem. There also exists a hybrid, site-bond percolation due to Heermann and Stauffer \cite{Heer}. In site percolation one assumes that the lattice sites are occupied at random with the probability $p$ (and hence with the probability $1-p$ are empty). A connected cluster is defined as a collection of all occupied sites that can communicate via the nearest-neighbor rule. In bond percolation, one thinks of clusters of connected conducting bonds instead. In this formulation all sites are initially occupied and bonds are occupied randomly with the probability $p$. Statistically, the $p$ value decides on how big the connected clusters could be for the given topology of the lattice. The key point of the theory is the existence of a critical value, the percolation threshold $p_c$ ($0 < p_c < 1$), above which the connected clusters span the entire lattice with the probability 1. The critical probability being smaller than 1 implies that the infinite clusters do not fill the ambient space yet. For $p < p_c$, the percolation dies away exponentially. The threshold value is non-universal: it depends on the type of the percolation problem (site, bond, or hybrid); details of the lattice (cubic, diamond, triangle, etc.); as well as the ambient dimensionality $d\geq 1$. The typical realizations of site and bond clusters on a square lattice are illustrated in Fig.~1. It is worth remarking that any point belonging to the infinite percolation cluster can be connected to the infinitely remote point via a connected escape path which lies everywhere on the cluster. For comprehensive reviews on percolation see, e.g., Refs. \cite{Isi,Stauffer,Naka,Havlin}. 

%
%
%
%
\begin{figure}
\includegraphics[width=1.00\textwidth]{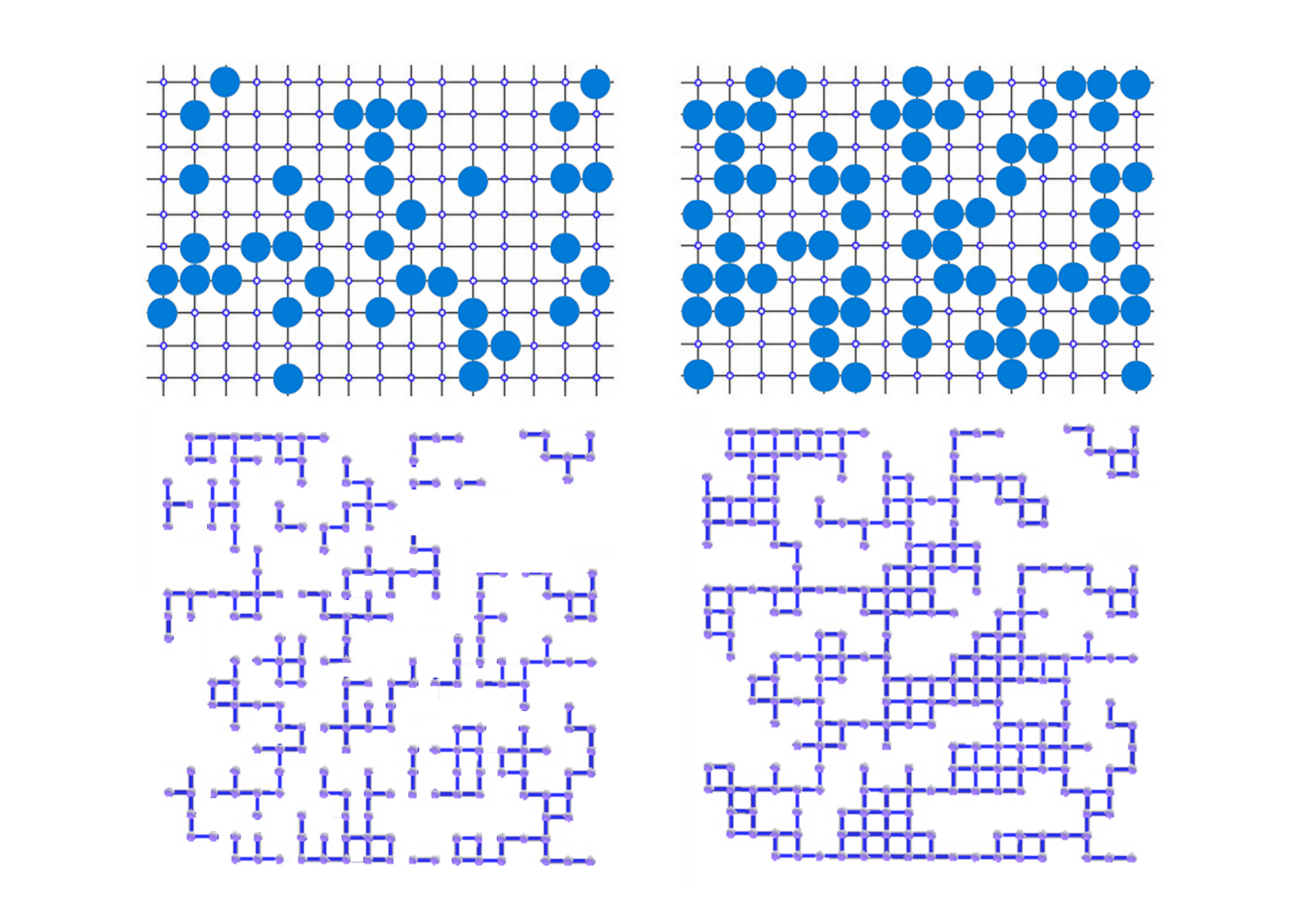}
\caption{\label{} Site {vs} bond percolation. Top: Site percolation problem, with circles representing occupied sites on a lattice. Left: A lattice system below the percolation threshold. Right: The same lattice system at the threshold of percolation, with a noticeably denser concentration of the occupied sites. Bottom: A similar picture for bond percolation. Connected (conducting) bonds are shown in blue color. Left: An insulating state of the lattice below the percolation threshold. Right: Random distribution of conducting bonds at the threshold of conducting dc electricity.}
\end{figure}

\subparagraph{Percolation critical exponents $\beta$, $\nu$, and $\mu$} Likewise to traditional critical phenomena, characterized by a scale-free statistics of the spontaneously occurring quantities, the geometry of connected clusters in vicinity of the percolation threshold is self-similar (fractal) \cite{Stauffer,Feder}. As $p\rightarrow p_c$, the percolation correlation (i.e., pair connectedness) length diverges as $\xi \propto |p-p_c|^{-\nu}$. For $p>p_c$, the probability to belong to the infinite cluster is $P_\infty (p) \propto (p-p_c)^{\beta} \propto \xi^{-\beta/\nu}$, whereas the dc conductivity behaves as $\sigma_{\rm dc} \propto (p-p_c)^{\mu} \propto \xi^{-\mu/\nu}$. The critical exponents $\beta$, $\nu$, and $\mu$ are universal in that they do not depend on the type of the percolation problem, nor on details of the lattice. They do depend on the ambient dimensionality $d$, however, and their numerical values are known in all $d\geq 1$. The ``mass" of a connected cluster scales with its size as $M(\xi) \propto \xi^d P_\infty (p) \propto \xi^{d-\beta/\nu}$, leading to a nontrivial Hausdorff dimension $d_f = d-\beta/\nu$. The latter expression is sometimes said to be ``hyperuniversal" as it holds in all $d\geq 1$. 

\subparagraph{Random walks on percolating clusters} The problem of diffusion on fractals \cite{Gennes,Straley,Gefen} has stirred considerable attention in the literature, especially, in terms of the random walk approach. If the random walker (an unbiased ``ant") is put on a connected cluster at percolation, then the distance it travels after $t$ time steps behaves as \cite{Havlin,Gefen} 
\begin{equation}
\left<{\bf {r}}^2 (t)\right> \propto t^{2/(2+\theta)}.\label{RW} 
\end{equation} 
The exponent $\theta$ is given by $\theta = (\mu - \beta) / \nu$. Remark that the dependence here is no longer proportional to the time $t$, by contrast with uniform spaces. Thus, diffusion is {\it anomalous}. The exponent $\theta$ describes topological characteristics of the fractal (such as connectivity, etc.) As such, it shows interesting invariance properties under smooth (diffeomorphic) maps of fractals \cite{UFN}. It has two common names in the literature: the connectivity exponent and the index of anomalous diffusion. This ambiguity merely reflects that the diffusion is anomalous because fractals possess anomalous connectedness features as voids are present at all scales. When $\theta\rightarrow 0$, normal (Fickian) diffusion is reinstalled. 

In a basic theory of percolation \cite{Stauffer} it is shown that $\mu > \beta$ for connected clusters, implying that $\theta > 0$. One sees that the mean-square displacement in Eq.~(\ref{RW}) grows slower-than-linear with time. This slowing down of the transport occurs as a result of multiple trappings and delays of the diffusing particles in cycles, bottlenecks, and deadends of the fractal object on which the random motions concentrate. Note that the scaling law above holds as a single-cluster rule (the ``ant" cannot jump between the clusters). Averaging over all clusters at percolation replaces Eq.~(\ref{RW}) by 
\begin{equation}
\left<{\bf {r}}^2 (t)\right> \propto t^{(2-\beta/\nu)/(2+\theta)}\label{RW+} 
\end{equation} 
for $t\ll \xi^{2+\theta}$. Equation~(\ref{RW+}) has implications for the ac conductivity at ``anomalous" frequency scales, $\omega\gg\xi^{-(2+\theta)}$, for which the charge carriers move only on the fractal \cite{Gefen,PRB01}. The various aspects of anomalous diffusion in fractal systems are summarized in the reviews \cite{Havlin,UFN,Bouchaud}.

\subparagraph{The spectral fractal dimension and the Alexander-Orbach conjecture} A hybrid parameter $d_s = 2d_f /(2+\theta)$ is often referred to as the spectral, or fracton, dimension. It is so called because it represents the density of states for vibrational excitations in fractal networks termed fractons \cite{Rammal,AO}. It also appears in the probability of the random walker to return to the origin ($\propto t^{-d_s / 2}$) \cite{Shaugh}. The key difference between the Hausdorff and spectral fractal dimensions lies in the fact that $d_f$ is a purely structural characteristic of the fractal, whereas $d_s$ mirrors the dynamical properties (via the connectedness issues) such as wave excitation, diffusion, etc. Note that the spectral fractal dimension is not larger than its Hausdorff counterpart, provided that the connectivity exponent $\theta \geq 0$. The value of $d_s$ can conveniently be considered as an effective fractional number of the degrees of freedom in fractal geometry, owing to the specific way it appears in the diffusion \cite{Gefen,Shaugh} and wave-propagation problems \cite{UFN,AO,OrbachD}. 

In the past years there has been much excitement about the Alexander-Orbach (AO) conjecture \cite{AO} that the spectral fractal dimension is exactly $4/3$ for percolation clusters in any ambient dimension $d$ greater than 1. This conjecture is important as it relates the structural characteristics of the fractal, contained in $d_f$, to the dynamical characteristics, contained in $\theta$. For $d\geq 6$, the AO conjecture was proven by Coniglio \cite{Coniglio} as a percolation problem on a Cayley tree (Bethe lattice). A Cayley tree is a graph without loops where each node contains the same number of branches (called the coordination number) \cite{Schroeder}. In many ways, owing to its hierarchical structure, a Cayley tree behaves as an infinite-dimensional space (its volume grows exponentially fast with the spatial scale). Not surprisingly, the percolation problem on a Cayley tree is regarded as a suitable model for mean-field percolation. For $d < 6$, the mean-field approach is invalidated as loops become important at all scales, thus impeding reduction to the trees. A great deal of effort has been invested to prove or disprove the AO conjecture in the lower embedding dimensions $d < 6$. It is now clear that in these dimensions the AO conjecture is not exact, nor does it generalize to all statistical fractals as for instance to the backbones of percolation clusters \cite{Havlin,StanCon}. Even so, the ``true" values found for the spectral fractal dimension at criticality continue to be numerically surprisingly close to the original AO result $d_s = 4/3$ for all $d\geq 2$ thus sustaining the conjecture. An interested reader may refer to the reviews \cite{Naka,Havlin}.  

\subparagraph{Percolation problem on the Riemann sphere} It is both interesting and instructive to demonstrate how the spectral fractal dimension may be obtained for threshold percolation on a plane ($d=2$). The main idea here \cite{PRE97} is to extend the plane by adding the point at infinity to it, then consider a stereographic projection of the infinite percolation cluster on the familiar from complex analysis Riemann sphere (see Fig.~2). Thus, the percolation problem will be compactified \cite{UFN}, as the point at infinity is mapped to the north pole.

\begin{figure}
\includegraphics[width=1.00\textwidth]{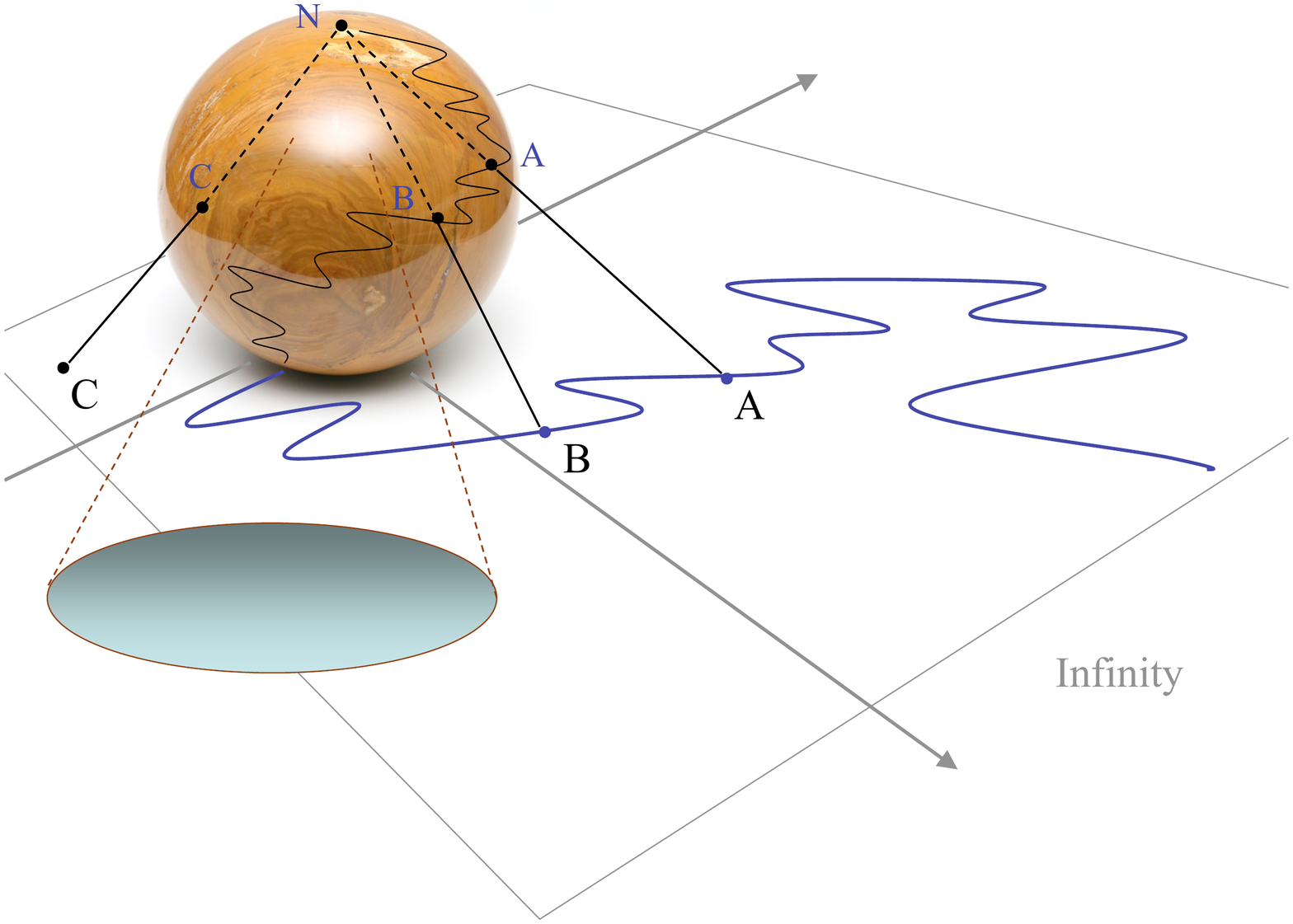}
\caption{\label{} Stereographic projection of a percolating cluster on the Riemann sphere. The north pole, $\rm N$, represents the point at infinity. When considered on the Riemann sphere a percolating escape path to infinity (blue line, with points $\rm A$ and $\rm B$ on it) originating from the south pole at which the sphere touches the plane is a simple arc connecting the two poles, south to north. A solid angle at the base of this arc has a lower bound, $\pi$, as dictated by connectedness. Then the mapping being conformal implies that at the threshold of percolation $\Omega_{d_s} = \pi$, leading to $d_s = 1.327\pm 0.001$.}
\end{figure}

Observe that the percolation cluster spanning the plane implies that its stereographic image covers part of the surface of the sphere, the north pole included. Without loss of generality, we may assume that the south pole at which the sphere touches the plane belongs to the cluster. When considered on the Riemann sphere a percolating escape path to infinity originating from the south pole will be a simple arc connecting the two poles, south to north. The key step is to notice that a solid angle at the base of this arc has a lower bound as posed by connectedness. Indeed this angle cannot be lesser than the angle at the base of half meridian. The latter is immediately seen to be equal to $\pi = \frac{1}{2}2\pi$. Clearly, the percolation cluster itself is based on a solid angle not lesser than this. It is convenient to think of the number $d_s$ of the degrees of freedom as corresponding to an orthogonal basis of $d_s$ vectors \cite{PRE97}, which span a fractional solid angle $\Omega_{d_s} = d_s \pi^{d_s /2}/\Gamma (d_s/2 +1)$. Here, $\Gamma$ denotes Euler's gamma function. Partial cases of this expression are, $\Omega_2 = 2\pi$ for $d_s = 2$ and $\Omega_3 = 4\pi$ for $d_s = 3$. Thus, we expect that, for connected clusters, $\Omega_{d_s} \geq \pi$, from which a lower bound on $d_s$ may be deduced by defining $\Omega_{d_s} = \pi$. We associate this lower bound with the threshold of macroscopic connectedness (threshold of percolation). In equating $\Omega_{d_s}$ to $\pi$ we used that the stereographic projection being a conformal map is angle and circle preserving. Putting all the various pieces together, we have \cite{UFN,PRE97}    
\begin{equation}
d_s \frac{\pi^{d_s /2}}{\Gamma (d_s/2 +1)} = 2 \frac{\pi^{d_s /2}}{\Gamma (d_s/2)} = \pi. \label{Pi} 
\end{equation}
Numerical solution shows that $d_s = 1.327\pm 0.001$, remarkably close to, although slightly smaller than, 4/3. Rigorously speaking, this result disproves the AO conjecture in $d=2$. Despite being this subtle, the observed deviation from $4/3$ is important as it helps avoid the secular terms problem when applying a renormalization-group technique near the percolation point \cite{Naka,Havlin}. Note that the solution to Eq.~(\ref{Pi}) has a remarkable meaning. It defines fractional dimensionality of a ball-like space seen from its center under the solid angle $\pi$. It is the dimensionality of this space, $d_s\approx 1.327$, which permits percolation in terms of a connected escape path to infinity. One sees that the percolation problem is essentially a topological problem. It decides on macroscopic connectedness of random systems in terms of the number of the coupled degrees of freedom. As such, the percolation problem has important implications for the dynamics of complex systems, and self-organized critical systems as particular case, as it will be demonstrated shortly.    

\subparagraph{Summary} Summarizing, the percolation problem presents a non-trivial problem with scale-free behavior. It is related with phase transition-like phenomena as well as the fundamental topology (via the connectedness issues). The percolation clusters provide a particularly clear example of statistical fractals in the limit $\xi\rightarrow\infty$. The percolation indices $\beta$, $\nu$, and $\mu$ are known in all ambient dimensions $d\geq 1$ and do not depend on details of the lattice nor the type of the percolation problem (universality). In addition to basic science, the percolation problem is of practical importance as it offers a platform for the description of transport properties of disordered (random) media. In particular, diffusion and electrical conduction problems on percolation clusters have been widely studied and discussed in the literature (Refs. \cite{Naka,Havlin,UFN}; references therein).  

\section{The SOC hypothesis}
\label{sec:2}

The challenge to understand fractals \cite{Mandel} and the $1/f$ ``noise" led Bak, Tang, and Wiesenfeld \cite{BTW} to introduce the concept of self-organized criticality, or SOC. The claim was that irreversible dynamics of systems with many coupled degrees of freedom (``complex" systems) would naturally generate self-organization into a critical state without fine tuning of any external or control parameter(s). By analogy with traditional critical phenomena it was argued that in vicinity of the critical state there is universal behavior, robust with respect to variations of parameters and with respect to randomness, and that the system can be characterized by power-laws and a set of critical exponents. An impressive list of publications\footnote{According to ISI's Web of Science the number of papers citing the seminal work in Ref. \cite{BTW} is above three thousand.} have been produced in the attempt to prove or disprove the SOC hypothesis for the various systems. The phenomenon was demonstrated on a number of automated lattice models, or ``sandpiles," displaying avalanche dynamics and scale invariance \cite{BTW,Tang,Zhang,Kadanoff}. The various aspects of self-organized criticality dynamics have been reviewed by Bak \cite{Work}, Jensen \cite{Jensen}, Turcotte \cite{Turcotte}, Charbonneau {\it et al.} \cite{Char} and Aschwanden \cite{Ash11}. 

To qualify as SOC, the system must be open, be coupled with the exterior, and involve many interacting degrees of freedom. In addition, its dynamics must be thresholded and nonlinear, and the driving, or energy injection, rate must be very slow (infinitesimal). An important advance of SOC is the realization that fractals appear naturally through a self-organization process and that the corresponding critical state is an attractor for the dynamics. In many ways the notion of SOC can be thought of as belonging to the nascent ``science of complex systems" which addresses the commonalities between apparently dissimilar natural, technological, and socio-economic phenomena, e.g., market crashes and climate disruptions \cite{Climate}. Despite its promising performance the SOC hypothesis is a subject of strong debate in the literature, and many issues related to it remain controversial or in demand for further investigation.   

\subparagraph{SOC vs percolation} Before we proceed with the main topics of this chapter, we would like to address the SOC hypothesis against the percolation problem discussed above. Indeed SOC shares with percolation the implications of threshold behavior and spatial self-similarity. An essential difference is that percolation is a purely geometrical model, whereas SOC involves, in addition, the temporal counterpart of the fractal, the $1/f$ noise \cite{1F}. In many ways SOC is a {\it spatio-temporal} phenomenon where both spatial and temporal self-similarities are coupled and long-ranged. 

Another important aspect is that in percolation and other traditional critical phenomena, control parameters must be fine tuned to obtain criticality (thus the name ``control"). In SOC phenomena, control parameters make part of the dynamical system instead: their values are defined dynamically as the system self-adjusts to accommodate the changing exterior conditions. It is in this sense that a SOC system is said to ``automatically" (without a fine tuning of parameters) reach the critical state.\footnote{Often one says that a SOC system possesses no tunable control parameters, but that's all about the wording.} More so, we remark that there exists a ``self-organized" formulation of some standard percolation processes such as the spread of deceases or forest fires. It was argued that their dynamics could be formulated so that they mimic SOC phenomena \cite{Grass}. The characteristic feature here is that singularities at $p_c$ emerge not in distribution of order parameters but of control parameters, making these phenomena look like SOC. This formulation is advantageous, as it leads to efficient numerical algorithms, allowing for a precise determination of the critical behavior, as for instance in models of self-organized critical directed percolation, with time interpreted as the preferred dimension \cite{Maslov}. 

\subparagraph{The ``guiding" mechanisms} An important issue concerns the mechanisms that ``guide" a system to criticality. These mechanisms are of two types. One type is associated with the application of an extremal principle that the dynamics should obey in order to satisfy the microscopic equations of motion. Examples of this type are the invasion percolation \cite{Invasion} and Bak-Sneppen models \cite{Sneppen,Quench}. In invasion percolation\footnote{To be distinguished from ordinary percolation discussed above.} $-$ introduced in physics by Wilkinson and Willemsen $-$ the dynamics proceed along a path of least resistance under the action of capillary forces. Under the condition that the flow rate is infinitesimal the system finds its critical points that are stable and self-organized \cite{Invasion}. The second type is associated with the operation of a nonlinear feedback between system's dynamical parameters \cite{PT} as for instance a feedback of the order parameter on the control parameter(s) as discussed by Sornette \cite{Sornette,Sor96}. The Bak, Tang, and Wiesenfeld's (BTW) sandpile is a prominent example of this type. In sandpiles the unstable sand slides off to decrease the slope and reinstall stability, thus providing a feedback of the particle loss process on the dynamical state of the pile. In many ways nonlinearity is an essential key element to SOC phenomena as it ensures a steady state where the system is marginally stable against a disturbance \cite{PT,Sornette}.           

\section{Going with the random walks: DPRW model}
\label{sec:3}

The percolation problem when account is taken for a dynamical feedback mechanism offers a suitable platform to build toy-models of self-organized critical phenomena. Early attempts in this direction refer to the ``dilution-by-hungry-ants" and the ``thermal-fuse" models (with and without a healing) \cite{Sornette}. In what follows, we discuss a model \cite{EPL,NJP}, dubbed dynamic polarization random walk (DPRW) model, which combines the implication of a feedback mechanism with the idea of random walks on a fractal cluster at percolation. The model is formulated as a transport problem for electrically charged particles of different kinds.\footnote{This electrical context of the model is non-crucial and can be relaxed \cite{NJP}.} The advent of random walks in place of automated lattice redistribution rules makes it possible to calculate the frequency-dependent complex susceptibility of the dynamical system at SOC along with the memory (response) function and in the end to obtain the SOC critical exponents in terms of three percolation critical indices $\beta$, $\mu$, and $\nu$. This approach paves the way for an analytical theory of SOC starting from the microscopic dynamical properties. One by-product of the random-walk model is a demonstration \cite{EPL,NJP} that the relaxation of a supercritical system to SOC is of Mittag-Leffler type \cite{Mittag-Leffler} (similarly to the relaxation in glassy systems and polymers). Indeed the Mittag-Leffler relaxation implies that behavior is multi-scale with a broad distribution of durations of relaxation events consistently with a description in terms of the fractional relaxation equation \cite{Klafter} and at odds with the well-known, Debye-like relaxation dynamics.

\subparagraph{Description of the model} We consider a hypercubic $d$-dimensional ($d\geq 1$) lattice confined between two opposite ($d-1$)-dimensional hyperplanes, which form a parallel-plate ``capacitor" as shown in Fig.~3. The plate on the right-hand-side is earthed. Free charges are built by external forces on the capacitor's left plate. When a unit free charge is added to the capacitor the lattice responds by burning a unit ``polarization" charge, which is an occupied site added at random to the lattice. When a unit free charge is removed from the capacitor a randomly chosen occupied site is converted into a ``hole" site (missing occupied site). A hole will be deleted from the system (converted into empty site) if/when the corresponding free charge has reached (or been moved to) the ground. There is a limit, $Q_{\max}$, on the amount of the free charges the capacitor can store, and this is defined as $Q_{\max} = e p_c N$, where $e$ is the elementary charge ($e=-1$), $p_c$ is the percolation threshold, and $N$ is the total number of sites across the lattice. Thus, the ability of the capacitor to store electric charges is limited to the occurrence of the infinite cluster at the percolation point. If, at any time, the above limit is exceeded, a double amount\footnote{This mimics non-zero inductance in the conduction process.} of the free charges in excess of $Q_{\max}$ will be removed from the capacitor and will be distributed between the sites of the infinite cluster with equal probability. The implication is that the capacitor leaks electric charges above the percolation point. This property reflects the onset of the dc conduction at the threshold of percolation.  

\begin{figure}
\includegraphics[width=1.00\textwidth]{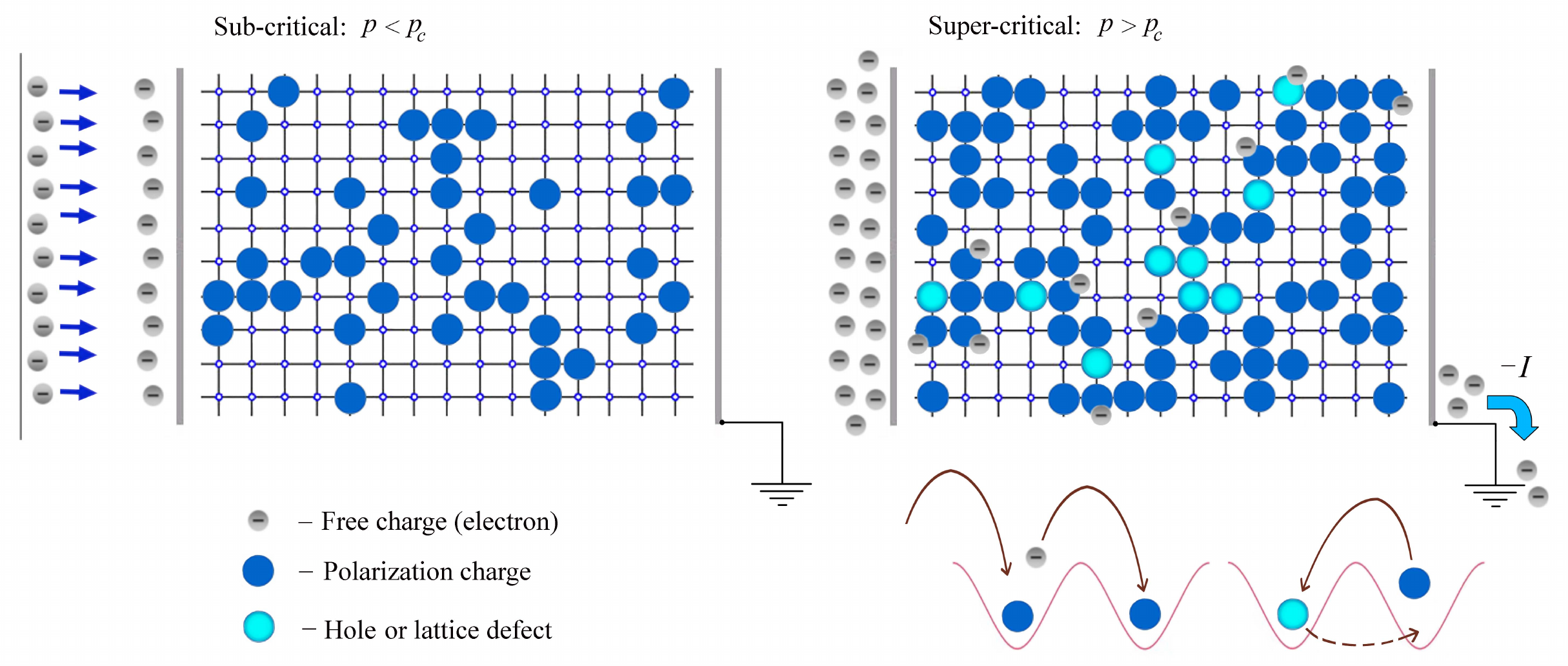}
\caption{\label{} Dynamic polarization random walk (DPRW) model. Grey, blue, and azure particles show respectively the free charges, polarization charges, and holes. Left: System below the percolation point. Free charges are built by external forces on the capacitor's left plate. Right: System slightly above the percolation point, with an illustration of hopping activities on the lattice. Adapted from Ref. \cite{NJP}.}
\end{figure}

When a hole appears on the infinite cluster it causes an activation event with the following consequence: One of the nearest-neighbor occupied sites, which is a random choice, will deliver its charge content to the hole. The hole which has just received the polarization charge becomes ordinary occupied site, while the donor site becomes a hole. The newborn hole, in its turn, will cause further activation event at the location where it has occurred thus triggering off a chain reaction of redistribution of polarization charges. The chain reaction continues until the hole reaches the earthed plate where it is absorbed (converted into empty site). When a hole appears on a finite cluster it causes a chain reaction of activation events in much a similar way as on the infinite cluster but with one modification regarding the ending of the activation: The chain reaction stops if (i) likewise to the infinite cluster case the hole reaches the earthed plate where it is converted into empty site, or if (ii) there are no more activities going on on the infinite cluster. In the latter case the finite cluster freezes in a ``glassy" state with the still holes in it until either a new hole appears on the infinite cluster or one or more occupied site are added to the lattice by external forces. 

\subparagraph{Random-walk hopping process} Essentially, the holes interchange their position with the nearest-neighbor occupied sites, and it appears reasonable to model this process as interchange hopping process \cite{Dyre}. In what follows we assume, following Ref. \cite{Gennes}, that there is a characteristic microscopic hopping time, which is taken to be unity, but more general hopping models can be obtained by introducing a distribution of waiting times between consecutive steps of the hopping motion (continuous time random walks, or CTRW's) \cite{CTRW,Wyss}. With the above assumption that the site acting as donor is a random choice the transport model is defined as random-walk hopping model. Similarly to the hole case, the free charges are assumed to behave as unbiased random walkers after their re-injection on the infinite cluster. They will hop at a constant rate between the nearest-neighbor occupied sites in random direction on the cluster on which they are initially placed until they reach the earthed plate where they get sink in the ground circuit (see Fig.~3). The holes act as the conducting sites for the motion of the free charges. The charged plate acts as a perfectly reflecting boundary. Hops to empty sites are forbidden. The latter condition limits the random walks to fractal geometry of the threshold percolation. 

\subparagraph{Dynamical geometry of threshold percolation} Overall, one can see that the system responds by chain reactions of random-walk hopping processes when it becomes slightly supercritical and it is quiescent otherwise. Excess free charges dissipating at the earthed plate provide a feedback mechanism by which the system returns to the percolation point. There will be a slowly (as compared to hopping motions) evolving dynamical geometry of the threshold percolation resulting from the competition between the adding of occupied sites to the lattice and the charge-releasing chain reactions. Based on the quantitative analysis below, we identify this state as a SOC state. This general picture based on the idea of a dynamic polarization response with random-walk hopping of the charge carriers might be called dynamic polarization random walk (DPRW) model.
 
In the DPRW SOC model, chain reactions of the hopping motion acquire the role of ``avalanches" in the traditional sandpiles. In the present analysis, we are interested in obtaining the critical exponents of the DPRW model by means of analytical theory. Numerical simulation of the DPRW dynamics is under way for comparison with the analytical predictions. By the time this chapter is being written, the characteristic signatures of multi-scale conductivity response of the dynamical system at criticality have been confirmed in the computer simulation model. In Fig.~4, we illustrate the existence of relaxation events of various sizes due to hole hopping on a 10$\times$10 square lattice with random distribution of the conducting nodes and the probability of site occupancy such as to mimic the percolation threshold and the conjectured SOC activities.

\begin{figure}
\includegraphics[width=1.00\textwidth]{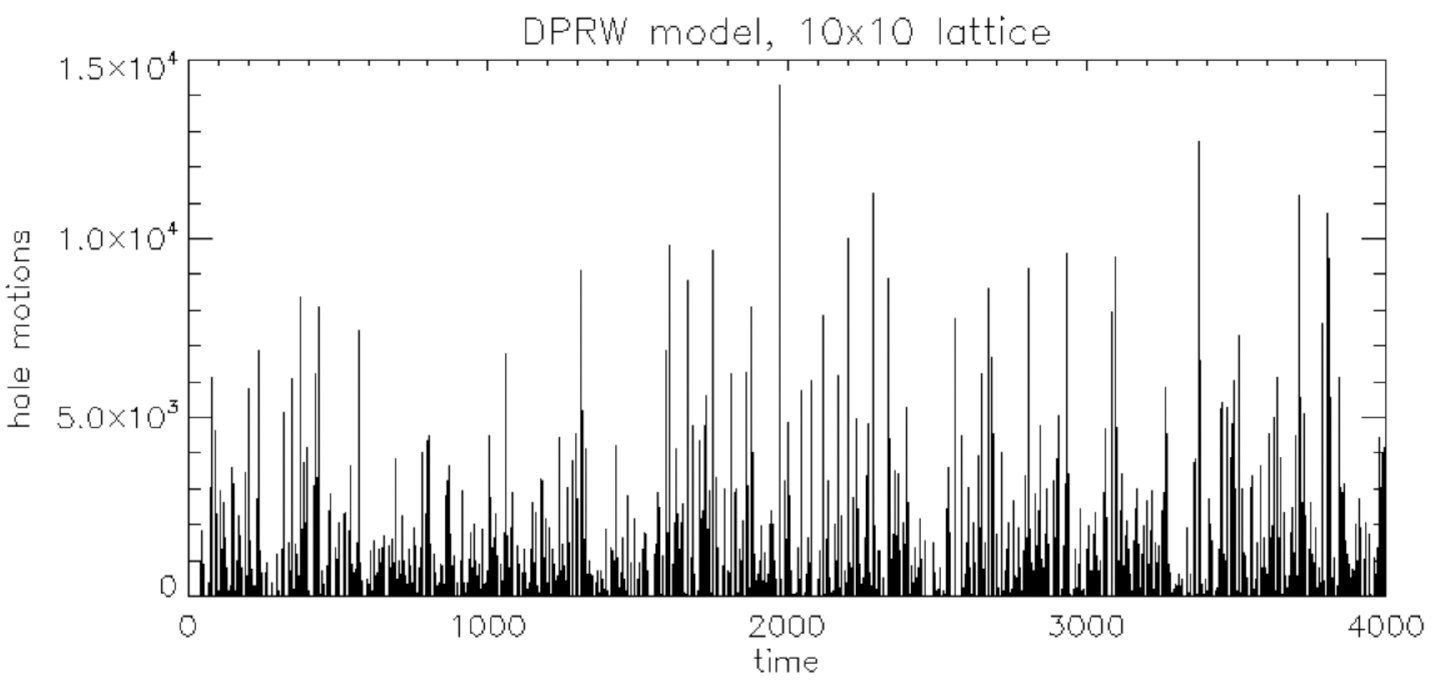}
\caption{\label{} Sizes of chain reactions due to hole hopping on a 10$\times$10 square lattice. Each spike corresponds to a chain reaction of hole hopping on the system-scale conducting cluster, with the height proportional to the number of hops to absorption at the earthed plate. This simple numerical realization illustrates the existence of relaxation events of various sizes consistently with the implication of SOC. Adapted from Ref. \cite{NJP}.}
\end{figure}

\section{Linear-response theory}

\subparagraph{Dynamics and orderings} Starting from an empty lattice (no potential difference between the plates), by randomly adding occupied sites to it, one builds the fractal geometry of the random, or uncorrelated, percolation, characterized by three percolation critical exponents $\beta$, $\nu$, and $\mu$ (connected clusters have fractal dimensionality $d_f = d - \beta / \nu$) \cite{Isi,Stauffer,Naka}. Remark that the infinite percolation cluster, in the true sense of the wording, exists only in the thermodynamic limit when the lattice itself is infinite. This limit arises because of the need to model the system-sized conducting clusters in terms of fractal geometry. In the absence of holes this percolation geometry is static (polarization charges can only move by exchanging their position with a hole) but when the holes appear on the lattice they cause local rearrangements in the distribution of the conducting sites. As a consequence, the conducting clusters on which the transport processes concentrate change their shape and their position in the real space. In the analysis of this section we shall require that the average number density of the holes be very small compared to the average number density of the polarization charges. The implication is that the system remains near the percolation point despite the slow evolution of the conducting clusters. Note that the lattice rules are such as to preserve the properties of the random percolation. In fact, no correlations are introduced in the distribution of the conducting sites at any step of the lattice update.  

\subparagraph{Frequency-dependent conductivity and diffusion coefficients} Given an input electric driving field ${\bf E} (t, {\bf r})$ the polarization response of the system is defined through 
\begin{equation}
{\bf P} (t, {\bf {r}}) = \int_{-\infty}^{+\infty} \chi (t - t^{\prime}) {\bf E} (t^{\prime}, {\bf {r}}) dt^{\prime},\label{PR} 
\end{equation}
where the response function $\chi (t - t^{\prime})$ is identically zero for $t < t^\prime$ as required by causality. We should stress that nonlocal integration over the space variable is not needed here in view of the local (nearest-neighbor) character of the lattice interactions. In a model in which the assumption of locality is relaxed as for instance in models permitting particle's L\'evy flights the integration over the space variable is expected to produce a physically nontrivial effect. We do not consider such models here. A Fourier transformed $\chi (t)$ defines the frequency-dependent complex susceptibility of the system, $\chi (\omega)$. In a basic theory of polarization response one also introduces the frequency-dependent complex ac conductivity, $\sigma_{\rm ac} (\omega)$, which is related to $\chi (\omega)$ by the Kramers-Kronig integral [Eq.~(\ref{KK}) below]. The dependence of the ac conductivity on frequency, specialized to the random walks on percolation systems \cite{Havlin,Gefen}, is given by the scaling relation $\sigma_{\rm ac} (\omega) \propto \omega ^\eta$, where the power exponent $\eta$ ($0\leq \eta\leq 1$) is expressible in terms of the percolation indices $\beta$, $\nu$, and $\mu$ as $\eta = \mu / (2\nu + \mu - \beta)$. We should stress that the scaling $\sigma_{\rm ac} (\omega) \propto \omega ^\eta$ incorporates conductivity responses from all clusters at percolation including those finite. In the DPRW model these implications are matched by the mechanism of the hole conduction permitting the polarization current on both infinite and finite clusters. The general linear-response theory expression \cite{Lax} for the conductivity $\sigma_{\rm ac} (\omega)$ in terms of the mean-square displacement from the origin $\left<{\bf {r}}^2 (t)\right>$ is
\begin{equation}
\sigma_{\rm ac} (\omega) = \frac{ne^2}{k_B T} D(\omega),\label{Ein} 
\end{equation}
where
\begin{equation}
\frac{1}{n_d} D(\omega) = \lim_{\epsilon\rightarrow 0+}\left[ (i\omega)^2 \int_0^\infty e^{-i\omega t} e^{-\epsilon t} \left<{\bf {r}}^2 (t)\right> dt\right] \label{Diff} 
\end{equation}
with $n_d$ a constant depending on the dimensionality of the lattice and $n$ and $e$ the density and charge of the carriers, respectively. The function $D(\omega)$ has the sense of the frequency-dependent diffusion coefficient \cite{Lax58}. In the zero-frequency limit, Eq.~(\ref{Ein}) reproduces the well-known Einstein relation between the static diffusion coefficient on the infinite cluster, $D_\infty$, and the dc conductivity, $\sigma_{\rm dc} = \lim_{\omega\rightarrow 0}\sigma_{\rm ac} (\omega)$. Note that the dc conductivity occurs only through the infinite cluster ($p > p_c$), as opposed to the ac conductivity response, which occurs through both finite and infinite clusters. In what follows, we require that the frequency $\omega$ be large compared to the characteristic evolution frequency in the distribution of the conducting sites. Denoting this last frequency by $\omega_*$, we have, for the present orderings, $\omega\gg\omega_*$. In this parameter range, $D(\omega) \propto \omega^\eta$. Consistently with the above definitions, the inverse frequency, $1/ \omega_*$, is ordered as the characteristic diffusion time on the infinite cluster, $\tau_* \simeq \xi^2 / D (\omega_*)$. Note that this time will depend on $\omega_*$ in accordance with Eq.~(\ref{Ein}). Hence, $\omega_* \simeq 1/\tau_* \simeq D (\omega_*) / \xi^2$, where $\xi \propto |p-p_c|^{-\nu}$ is the diverging percolation pair connectedness length; $p$ is the probability of site occupancy; and $p_c$ is the percolation threshold. We have, at the margins of self-similar behavior, $D (\omega_*) \propto (\omega_*) ^\eta$, implying that $\omega_* \propto |p-p_c|^{2\nu/(1-\eta)}$. Observe that $\omega_*\rightarrow 0$ for $p\rightarrow p_c$. Remembering that there is a microscopic hopping time, which is taken to be unity, we assess the Kubo number \cite{Kubo,Frisch} in vicinity of the SOC state as 
\begin{equation}
Q_* \simeq 1/\omega_*\xi \propto |p-p_c|^{-\nu (1+\eta)/(1-\eta)}. \label{Ku} 
\end{equation}
One sees that $Q_*\rightarrow\infty$ in the limit $p\rightarrow p_c$. The Kubo number is a suitable dimensionless parameter which quantifies how the evolution processes in the lattice compare with the microscopic hopping motions. The divergency of the Kubo number at criticality implies that there is a time scale separation: fast hopping motions vs infinitesimal evolution change. In terms of the Kubo number ($Q_*\rightarrow\infty$), the diffusion coefficient $D (\omega_*)$ becomes
\begin{equation}
D (\omega_*) \propto \omega_* Q_* ^{\gamma}, \label{PS} 
\end{equation}
where we have introduced $\gamma = 1-\eta$. This scaling law appears in models of anomalous diffusion by low-frequency turbulence \cite{PAppl,PAppl+,Zim01,PRE01,PRE09}. Special cases of Eq.~(\ref{PS}) include the well-known Bohm scaling \cite{Bohm}, characterized by $\gamma = 1$, as well as the anomalous so-called ``percolation" scaling ($\gamma\approx 0.7$), dating back to diffusion-advection models of Isichenko and co-workers \cite{Isi,Kalda,Kalda2}. Alternatively, the diffusion coefficient on a time varying fractal distribution, Eq.~(\ref{PS}), can be deduced from the general scaling form \cite{PRE09}  
\begin{equation}
\langle{\bf r}^2 (t)\rangle = \xi^{2} (t/\tau_*) f (t/\tau_*), \label{9ST} 
\end{equation}
where $f$ is a scaling function, which interpolates between the initial-time power-law and flat asymptotic ($t\rightarrow+\infty$) behavior: $f(\infty) = {\rm const.}$ The form in Eq.~(\ref{9ST}) is similar to that considered by Gefen {\it et al.} \cite{Gefen} for anomalous diffusion on percolation clusters (in their model, $\tau_*\propto\xi^{2+\theta}$), and earlier by Straley \cite{Straley}.  

\subparagraph{Power-law power spectral density} By applying the Kramers-Kronig relations ${\rm Im}\, \chi (\omega) \propto \sigma_{\rm ac} (\omega) / \omega$ and 
\begin{equation}
{\rm Re}\, \chi (\omega) \propto \mathrm{V.P.} \int \frac{d\omega^{\prime}}{\omega^{\prime} (\omega^{\prime} - \omega)} \sigma_{\rm ac} (\omega^{\prime}) \label{KK} 
\end{equation}
it is found that $\chi (\omega) \propto \omega ^{-\gamma}$, with $\gamma = 1-\eta$. A Fourier transformed Eq.~(\ref{PR}) reads ${\bf P} (\omega, {\bf {r}}) = \chi (\omega) {\bf E} (\omega, {\bf {r}})$. One can see that the power spectral density, $S (\omega)$, of the system response to a white-noise perturbation, ${\bf E} (\omega, {\bf {r}}) = \bf 1$, will be proportional to $|\chi (\omega)|^2$. The end result reads: 
\begin{equation}
S (\omega) \propto |\chi (\omega)|^2 \propto |\sigma_{\rm ac} (\omega) / \omega|^2 \propto \omega^{-\alpha}, \label{PSD} 
\end{equation}
where $\alpha = 2(1-\eta) = 2\gamma$. The conclusion is that the power spectral density in the DPRW model is given by an inverse power-law distribution, with the $\alpha$ value depending on scaling properties of the ac conductivity response. 

\subparagraph{Stretched-exponential relaxation and the distribution of relaxation times} Next, we obtain the distribution of relaxation times self-consistently. For this, assume that the system is slightly supercritical, then consider a charge density perturbation, $\delta\rho (t,{\bf r})$, caused by the presence of either free charges or holes on the conducting clusters. ``Slightly supercritical" means that the dependence of the ac conductivity response on frequency can, with good accuracy, be taken in the power-law form $\sigma_{\rm ac} (\omega) \propto \omega^{1-\gamma}$ discussed above. The implication is that at adding $\delta\rho (t,{\bf r})$ to the conducting system at percolation we neglect the departure of the systems geometric properties from pure self-similarity. Without loss in generality, we assume that the perturbation $\delta\rho (t,{\bf r})$ is created instantaneously at time $t=0$. That means that the function $\delta\rho (t,{\bf r})\equiv 0$ for $t < 0$ for all ${\bf r}$. The perturbation $\delta\rho (t,{\bf r})$ generates an electric field inhomogeneity, $\delta {\bf E} (t,{\bf r})$ in accordance with Maxwell's equation ${\bf \nabla}\cdot \delta {\bf E} (t,{\bf r}) = 4\pi\delta\rho (t,{\bf r})$. Consistently with the above discussion, we adopt that for $t > 0$ the decay of $\delta\rho (t,{\bf r})$ is due to the spreading of charge-carrying particles (electrons and/or holes) via the random walks on the underlying fractal distribution. The polarization response to $\delta {\bf E} (t,{\bf r})$ is given by
\begin{equation}
\delta {\bf P} (t, {\bf {r}}) = \int_{-\infty}^{+\infty} \chi (t - t^{\prime}) \delta {\bf E} (t^{\prime}, {\bf {r}}) dt^{\prime},\label{Delta} 
\end{equation}
where, as usual, $\chi (t - t^{\prime})\equiv 0$ for $t < t^{\prime}$. The density of relaxation currents is defined as the time derivative of $\delta {\bf P} (t, {\bf {r}})$, i.e.,  
\begin{equation}
\delta {\bf j} (t,{\bf r}) = \frac{\partial}{\partial t} \int_{-\infty}^{+\infty} \chi (t - t^{\prime}) \delta {\bf E} (t^{\prime}, {\bf {r}}) dt^{\prime}. \label{DRC} 
\end{equation}
The continuity implies that  
\begin{equation}
\frac{\partial}{\partial t}\delta\rho (t,{\bf r}) + {\bf \nabla}\cdot\frac{\partial}{\partial t} \int_{-\infty}^{+\infty} \chi (t - t^{\prime}) \delta {\bf E} (t^{\prime}, {\bf {r}}) dt^{\prime} = 0.\label{2} 
\end{equation}
Taking ${\bf \nabla}\cdot$ under the integral sign, then eliminating $\delta {\bf E} (t,{\bf r})$ by means of Maxwell's equation ${\bf \nabla}\cdot \delta {\bf E} (t,{\bf r}) = 4\pi\delta\rho (t,{\bf r})$, we find, with the self-consistent charge density,
\begin{equation}
\frac{\partial}{\partial t}\left(\delta\rho (t,{\bf r}) + 4\pi\int_{-\infty}^{+\infty} \chi (t - t^{\prime}) \delta\rho (t^{\prime}, {\bf {r}}) dt^{\prime}\right) = 0.\label{3} 
\end{equation}
In writing Eqs.~(\ref{2}) and~(\ref{3}) we have also assumed that $t > 0$. We now integrate in Eq.~(\ref{3}) to find
\begin{equation}
\delta\rho (t,{\bf r}) + 4\pi\int_{-\infty}^{+\infty} \chi (t - t^{\prime}) \delta\rho (t^{\prime}, {\bf {r}}) dt^{\prime} = g ({\bf r}).\label{3+} 
\end{equation}
Here, the function $g ({\bf r})$ is an arbitrary function of the position vector ${\bf r}$, which appears in the derivation as the constant of integration over time. Under the conditions $\chi (t - t^{\prime})\equiv 0$ for $t < t^{\prime}$ and $\delta\rho (t,{\bf r})\equiv 0$ for $t < 0$ for all ${\bf r}$, Eq.~(\ref{3+}) reduces to 
\begin{equation}
\delta\rho (t,{\bf r}) + 4\pi\int_{0}^{t} \chi (t - t^{\prime}) \delta\rho (t^{\prime}, {\bf {r}}) dt^{\prime} = g ({\bf r}).\label{3++} 
\end{equation}
If we allow $t\rightarrow +0$, we find that for $\gamma > 0$ the integral term on the left-hand-side goes to zero (as $\propto t^\gamma$):
\begin{equation}
\lim_{t\rightarrow +0} \int_{-\infty}^{+\infty} \chi (t - t^{\prime}) \delta\rho (t^{\prime}, {\bf {r}}) dt^{\prime} = \lim_{t\rightarrow +0} \int_{0}^{t} \chi (t - t^{\prime}) \delta\rho (t^{\prime}, {\bf {r}}) dt^{\prime} = 0,\label{3+++} 
\end{equation}
from which it is clear that $g ({\bf r}) = \lim_{t\rightarrow +0} \delta\rho (t,{\bf r})$. We consider this last condition as the initial condition for the relaxation problem. Essentially the same condition holds in the limit $\gamma \rightarrow 0$, provided that $\lim_{t\rightarrow +0}$ is taken first. A Fourier transformed Eq.~(\ref{3++}) reads 
\begin{equation}
\delta\rho (\omega,{\bf k}) + 4\pi \chi (\omega) \delta\rho (\omega, {\bf {k}}) = g ({\bf k}) / \omega,\label{FT} 
\end{equation}
where ${\bf k}$ is position vector in reciprocal space, and $g ({\bf k})$ is the Fourier image of $g ({\bf r})$. Writing the susceptibility as $\chi (\omega) = \tau_\lambda^{-\gamma} \omega^{-\gamma} / 4\pi$ with $\tau_\lambda$ a time constant it is found that 
\begin{equation}
\delta\rho (\omega,{\bf k}) = \frac{1}{\omega + \tau_\lambda^{-\gamma} \omega^{1-\gamma}} g ({\bf k}).\label{DD} 
\end{equation}
The quantity $\tau_\lambda$ has the sense of lifetime of a perturbation with wavelength $\lambda$. We expect that $\tau_\lambda \propto\lambda^{z}$ at criticality, where $z$ is a scaling exponent. A derivation of this scaling relation will be given shortly. Separating the variables, we write $\delta\rho (\omega,{\bf k}) = \varphi (\omega) g ({\bf k})$, with  
\begin{equation}
\varphi (\omega) = 1/(\omega+ \tau_\lambda^{-\gamma} \omega^{1-\gamma}), \label{Inv} 
\end{equation}
which we consider as the relaxation function in the frequency domain. On inversion to the time domain, Eq.~(\ref{Inv}) generates the Mittag-Leffler function, $E_\gamma [-(t/\tau_\lambda)^\gamma]$, which has series expansion \cite{Mittag-Leffler,Klafter}
\begin{equation}
E_\gamma [-(t/\tau_\lambda)^\gamma] = \sum _{n=0}^\infty \frac{[-(t/\tau_\lambda)^\gamma]^n}{\Gamma (1+\gamma n)}. \label{ML} 
\end{equation}
Thus, $\varphi (t) = E_\gamma [-(t/\tau_\lambda)^\gamma]$. One sees that the relaxation to SOC of a supercritical state is described by the Mittag-Leffler function $E_\gamma [-(t/\tau_\lambda)^\gamma]$, and not by a simple exponential function as for standard relaxation. We note in passing that the Mittag-Leffler function is the natural generalization of the exponential function. The latter is included as a special case $\gamma = 1$. For times $t \ll \tau_\lambda$, the Mittag-Leffler function, Eq.~(\ref{ML}), can be approximated by a stretched-exponential the so-called Kohlrausch-Williams-Watts (KWW) relaxation function \cite{Kohlrausch,Watts}
\begin{equation}
E_\gamma [-(t/\tau_\lambda)^\gamma]\simeq \exp [- (t/\tau_\lambda) ^{\gamma} / \Gamma (1+\gamma)],\label{KWW} 
\end{equation} 
which is often found empirically in various amorphous materials as for instance in many polymers and glass-like materials near the glass transition temperature (for reviews see Refs. \cite{Phillips} and \cite{Kaatz}, and references therein). The KWW relaxation function can conveniently be considered \cite{Montroll} as a weighted average of the ordinary exponential functions, each corresponding to a single relaxation event in the system:
\begin{equation}
\exp [- (t/\tau_\lambda) ^{\gamma} / \Gamma (1+\gamma)] = \int_0^{\infty} e^{-t/\Delta t} w_{\gamma} (\Delta t) d\Delta t.\label{Mont} 
\end{equation}
The weighting function $w_{\gamma} (\Delta t)$ is given by Eqs. (51d) and (55) of Ref. \cite{Montroll} where one replaces the exponent $\alpha$ with $\gamma$, the time constant $T$ with $\tau_\lambda$, and the variable $\mu$ with $\tau_\lambda / \Delta t$. In our notations:
\begin{equation}
w_{\gamma} (\Delta t) = (\tau_\lambda / \Delta t^2) L_{\gamma, -1} (\tau_\lambda / \Delta t),\label{Skew} 
\end{equation}
where $L_{\gamma, -1}$ is the L\'evy distribution function with skewness $-1$ (e.g., Ref. \cite{Wolf}). Assuming a long-wavelength perturbation (i.e., the parameter $\lambda$ being much longer than the microscopic lattice distance: $\lambda\gg 1$), and setting $\tau_\lambda/\Delta t \gg 1$, we can further approximate the L\'evy distribution $L_{\gamma, -1}$ by the Pareto inverse-power distribution. This gives $L_{\gamma, -1} (\tau_\lambda / \Delta t) \propto (\tau_\lambda / \Delta t)^{-(1+\gamma)}$ leading to a pure power-law distribution of relaxation times, consistently with the wisdom of SOC: 
\begin{equation}
w_{\gamma} (\Delta t) \propto \Delta t ^{-2}\Delta t^{1+\gamma}\propto \Delta t ^{-\eta}. \label{Skew2} 
\end{equation}
These distributions were earlier conjectured for SOC \cite{Tang}. Our conclusion so far is that the relaxations are multi-scale, in accordance with Eq.~(\ref{Mont}), and their durations are power-law distributed. The distribution is heavy-tailed in the sense that $\int d\Delta t w_{\gamma} (\Delta t)\propto \tau_\lambda^{\gamma}\rightarrow\infty$ for $\tau_\lambda\rightarrow\infty$.    

\subparagraph{Consistency check} In a basic theory of dielectric relaxation one writes the frequency-dependent complex dielectric parameter as \cite{Montroll,Constant}
\begin{equation}
\epsilon (\omega) - 1 \propto -\int _0^\infty \frac{d\varphi (t)}{dt} e^ {i\omega t} dt, \label{DieCo} 
\end{equation}
where $\varphi (t)$ is the relaxation function that describes the decay of polarization after the polarizing electric field has been stepped down or removed instantaneously. In the DPRW model, a step-down type electric field occurs as a consequence of re-injection of the free charges to the infinite cluster. The ensuing relaxation dynamics are mimicked by the chain reactions of hole hopping which act as to properly redistribute the polarization charges across the lattice. Based on the above analysis, we identify the relaxation function in Eq.~(\ref{DieCo}) with the Mittag-Leffler function to yield $\varphi (t) \simeq E_\gamma [-(t/\tau_\lambda)^\gamma]$. In vicinity of the critical state, because the upper limit on $\tau_\lambda$ diverges, we can, moreover, replace $E_\gamma [-(t/\tau_\lambda)^\gamma]$ with $\exp [- (t/\tau_\lambda) ^{\gamma} / \Gamma (1+\gamma)]$ for (almost) all $0 < t \leq\infty$. Thus, for $p\rightarrow p_c$, $\varphi (t) \simeq \exp [- (t/\tau_\lambda) ^{\gamma} / \Gamma (1+\gamma)]$. Integrating by parts in Eq.~(\ref{DieCo}), after a simple algebra one obtains
\begin{equation}
\epsilon (\omega) - 1 \propto 1 - s \mathrm{V} (s) + is \mathrm{Q} (s), \label{DieCo2} 
\end{equation}
where $s = \omega \tau_\lambda$ is dimensionless frequency, and ${\rm Q} (s)$ and ${\rm V} (s)$ are the L\`evy definite integrals:
\begin{equation}
\mathrm{Q} (s) = \int_0^{+\infty}\exp{(-u^{\gamma})} \cos{(us)}du, \label{QQ}
\end{equation}
\begin{equation}
\mathrm{V} (s) = \int_0^{+\infty}\exp{(-u^{\gamma})} \sin{(us)}du. \label{VV}
\end{equation}
In the parameter range of multi-scale relaxation response, $\tau_\lambda/\Delta t \gg 1$, $\omega \tau_\lambda \gg 1$, the following series expansions of the L\`evy integrals hold \cite{Montroll}: 
\begin{equation}
\mathrm{Q} (s) = \sum _{n=1}^\infty (-1)^{n-1} \frac{1}{s^{n\gamma + 1}} \frac{\Gamma (n\gamma + 1)}{\Gamma (n + 1)} \sin \frac{n\gamma\pi}{2},\label{SQQ}
\end{equation} 
\begin{equation}
\mathrm{V} (s) = \sum _{n=0}^\infty (-1)^{n} \frac{1}{s^{n\gamma + 1}} \frac{\Gamma (n\gamma + 1)}{\Gamma (n + 1)} \cos \frac{n\gamma\pi}{2}.\label{SVV}
\end{equation}    
From Eqs.~(\ref{SQQ}) and~(\ref{SVV}) one can see that the expansion of $\mathrm{Q} (s)$ starts from a term which is proportional to $s^{-(1+\gamma)}$, and so does the expansion of $\mathrm{V} (s) - 1/s$. Hence, up to higher order terms, $\epsilon (\omega) - 1 \propto s^{-\gamma}$. Given this, one applies the Kramers-Kronig relations $s \mathrm{Q} (s) \propto \sigma_{\rm ac} (\omega) / \omega$ and 
\begin{equation}
1 - s \mathrm{V} (s) \propto \mathrm{V.P.} \int \frac{d\omega^{\prime}}{\omega^{\prime} (\omega^{\prime} - \omega)} \sigma_{\rm ac} (\omega^{\prime}) \label{KK2} 
\end{equation}
to find the scaling of the ac conduction coefficient to be $\sigma_{\rm ac} (\omega) \propto \omega ^{1-\gamma}$. By comparing this with the above expression $\sigma_{\rm ac} (\omega) \propto \omega ^{\eta}$ one reiterates that $\gamma = 1 - \eta$ consistently with the distribution of durations of relaxation events, Eq.~(\ref{Skew2}). 

\subparagraph{Fractional relaxation and diffusion equations} As was shown by Gl\"ockle and Nonnenmacher \cite{Nonn}, the Mittag-Leffler function, Eq.~(\ref{ML}), is the solution of the fractional relaxation equation 
\begin{equation}
\tau_\lambda^{\gamma}\frac{d\varphi (t)}{dt} = - {_0}D_t^{1-\gamma} \varphi (t), \label{FRE} 
\end{equation} 
where
\begin{equation}
{_0}D_t^{1-\gamma} \varphi (t) = \frac{1}{\Gamma (\gamma)}\frac{\partial}{\partial t}\int _{0}^{t} \frac{\varphi (t^{\prime})}{(t - t^{\prime})^{1-\gamma}}dt^\prime \label{23} 
\end{equation}
is a fractional time the so-called Riemann-Liouville derivative \cite{Klafter,Podlubny}. Partial cases of this derivative are the unity operator for $\gamma\rightarrow 1$ and $\partial /\partial t$ for $\gamma\rightarrow 0$. It is noticed, following Ref. \cite{Sokolov}, that the Mittag-Leffler function $E_\gamma [-(t/\tau_\lambda)^\gamma]$ describes the relaxation toward equilibrium of particles governed by the fractional diffusion equation, or FDE
\begin{equation}
\frac{\partial}{\partial t} P (t, {\bf{r}}) = {_0}D_t^{1-\gamma} {\bf {\nabla}}^2 P (t, {\bf{r}}),\label{24} 
\end{equation} 
where $P (t, {\bf{r}})$ is the probability density of finding a particle (random walker) at time $t$ at point ${\bf{r}}$, and the Laplacian operator stands for the local (nearest-neighbor) character of the lattice interactions. 

\subparagraph{Derivation of the fractional diffusion equation} It is instructive to obtain the fractional diffusion equation, Eq.~(\ref{24}), directly from the DPRW relaxation model. For this, let us introduce the electrostatic potential, $\delta\Phi (t,{\bf r})$, corresponding to the electric field inhomogeneity, $\delta {\bf E} (t,{\bf r}) = -\nabla\delta\Phi (t,{\bf r})$. Upon substituted into Eq.~(\ref{2}), 
\begin{equation}
\frac{\partial}{\partial t}\delta\rho (t,{\bf r}) = \frac{\partial}{\partial t} \int_{-\infty}^{+\infty} \chi (t - t^{\prime}) \nabla^2 \delta\Phi (t^{\prime}, {\bf {r}}) dt^{\prime}.\label{2phi} 
\end{equation}
In the vicinity of self-organized critical state, we can represent the total charge density, $\rho (t,{\bf r})$, as a sum of ``unperturbed" or background density, $\rho_c = |e| p_c$, and a perturbation, $\delta\rho (t,{\bf r})$, describing the deviation from criticality: $\rho (t,{\bf r}) = \rho_c + \delta\rho (t,{\bf r})$. To obtain the dependence of $\rho (t,{\bf r})$, we use the effective-medium approximation \cite{Effective}, a standard technique for calculating average physical properties of many-body systems. The idea is to think of the particles as embedded into an ``effective" potential, $\delta\Phi (t,{\bf r})$, where they will be Boltzmann-distributed in accordance with   
\begin{equation}
\rho (t,{\bf r}) = \rho_c \exp \left[e\delta\Phi (t, {\bf {r}}) / T\right].\label{Boltz} 
\end{equation}
Self-consistently, one requires that, on the average, the embedding in the effective medium has the same overall property as the effective medium itself \cite{Dyre,Effective}. In writing Eq.~(\ref{Boltz}) we took into account that the perturbation, $\delta\rho (t,{\bf r})$, is due to negatively charged particles (electrons and/or holes). The normalization condition is defined through $\rho (t,{\bf r}) \rightarrow \rho_c$ for $\delta\Phi (t, {\bf {r}}) \rightarrow 0$. In the above, the parameter $T$ has the sense of ``thermodynamic temperature," associated with the random motion of particles on a conducting cluster. For $|e\delta\Phi (t, {\bf {r}})| / T \ll 1$, expanding the exponential function on the right of Eq.~(\ref{Boltz}), one finds $\delta\rho (t,{\bf r}) \approx (p_c e^2 / T) \delta\Phi (t, {\bf {r}})$. Thus, for small perturbations (high temperatures), $\delta\rho (t,{\bf r})$ is proportional to $\delta\Phi (t,{\bf r})$, as it should. Eliminating $\delta\Phi (t, {\bf {r}})$ in Eq.~(\ref{2phi}), we have
\begin{equation}
\frac{\partial}{\partial t}\delta\rho (t,{\bf r}) = \frac{\partial}{\partial t} \int_{-\infty}^{+\infty} \chi (t - t^{\prime}) (T / p_c e^2) \nabla^2 \delta\rho (t^{\prime}, {\bf {r}}) dt^{\prime}.\label{2rho} 
\end{equation} 
The memory function, $\chi (t)$, is obtained as Fourier inversion of $\chi (\omega) = \tau_\lambda^{-\gamma} \omega^{-\gamma} / 4\pi$, yielding $\chi (t) \propto t^{\gamma - 1}$. Under the conditions $\chi (t - t^{\prime})\equiv 0$ for $t < t^{\prime}$ and $\delta\rho (t,{\bf r})\equiv 0$ for $t < 0$ for all ${\bf r}$, the improper integration in Eq.~(\ref{2rho}) can be performed in the limits from 0 to $t$. Collecting all dimensional and numerical parameters in one effective ``diffusion coefficient," $A_\gamma \propto (T / 4\pi p_c e^2) \tau_\lambda^{-\gamma}$, we can write  
\begin{equation}
\frac{\partial}{\partial t}\delta\rho (t,{\bf r}) = \frac{1}{\Gamma (\gamma)}\frac{\partial}{\partial t} \int_{0}^{t} \frac{A_\gamma}{(t-t^{\prime})^{1-\gamma}} \nabla^2 \delta\rho (t^{\prime}, {\bf {r}}) dt^{\prime},\label{End} 
\end{equation} 
which is an equivalent form of Eq.~(\ref{24}). The fractional diffusion equation, Eq.~(\ref{End}), can be thought of as deriving from the generalized ``Fick's law"  
\begin{equation}
\delta {\bf j} (t,{\bf r}) = - \frac{1}{\Gamma (\gamma)}\frac{\partial}{\partial t} \int_{0}^{t} \frac{A_\gamma}{(t-t^{\prime})^{1-\gamma}} \nabla \delta\rho (t^{\prime}, {\bf {r}}) dt^{\prime}, \label{Fick} 
\end{equation}
or $\delta {\bf j} (t,{\bf r}) = - {_0}D_t^{1-\gamma}A_\gamma\nabla \delta\rho (t, {\bf {r}})$, which is readily deduced from Eq.~(\ref{DRC}) in the effective-medium approximation. Equation~(\ref{Fick}) can equivalently be obtained \cite{PRE09} from the general scaling law for anomalous diffusion on percolation systems. A derivation using the scheme of continuous time random walks (CTRW's) can be found in Refs. \cite{Ctrw1,Ctrw2}. We should stress that the non-Markovian nature of Eqs.~(\ref{End}) and~(\ref{Fick}), together with the fractional relaxation equation, Eq.~(\ref{FRE}), accounts for the long-time memory effects in SOC phenomena, where one believes \cite{Sornette} the time evolution exhibits long tails and infinite correlation scale. 

The occurrence of the fractional diffusion equation, Eq.~(\ref{End}), might be interpreted, with the aid of the proposed SOC model, in favor of considering SOC as one important case for {\it fractional kinetics} \cite{Nature}. The concept of fractional kinetics enters different areas of research, such as turbulent transport in plasmas and fluids, particle dynamics in potential fields, quantum optics, and many others. This subject is summarized in comprehensive reviews \cite{Klafter,Report,Klafter2}. In many ways equations built on fractional derivatives offer an elegant and powerful tool to describe anomalous transport in complex systems \cite{Report}. There is an insightful connection with a generalized master equation formalism along with a mathematically convenient way for calculating transport moments as well as solving initial and boundary value problems \cite{Klafter,Klafter2}. The fundamental solution or Green's function of the fractional Eq.~(\ref{24}) is evidenced in Table~1 of Ref.~\cite{Klafter2}. 

\subparagraph{Dispersion-relation exponent, Hurst exponent, and the $\tau$-exponent} In sandpile SOC models, one is interested in how the lifetime of an activation cluster scales with its size \cite{Zhang}. In the DPRW model by activation cluster one means a connected cluster of activated sites. An occupied site is said ``activated" if it has become a hole or if it contains a free charge. Clearly, activation clusters can only exist above the percolation threshold. Note that activation clusters are subsets of the underlying conducting cluster of polarization charges. The notion of activation cluster is but a visualization of the charge density inhomogeneity $\delta\rho (t,{\bf r})$ in terms of a connected distribution of activated sites. Activation clusters decay because the constituent charged particles (holes and/or free charges) diffuse away via the random walks. 

Consider an isotropic activation cluster composed of the free particles. (The nature of the particles does not matter here $-$ the hole case is just similar.) It is assumed for convenience, without loss of generality, that each site of the activation cluster contains only one particle. Thus, the number density of the free particles inside the activation area is equal to one. It steps down to zero just outside. If the microscopic lattice distance is $a$ ($a=1$), then there is a unit density gradient across the boundary of the activation cluster looking inside. Because of this gradient the activation cluster will be loosing particles on the average. A particle that has crossed the boundary against the direction of the gradient is considered lost from the cluster. As the particles dissipate, the location of the density pedestal shifts inward with speed $u$. The local flux density of those particles leaving per second the activation area is just the gradient times the local diffusion coefficient. The latter depends on frequency of the relaxation process as $D(\omega) \propto \omega ^\eta$ in accordance with Eq.~(\ref{Ein}). If $l$ is the current size of the cluster, then the corresponding relaxation frequency is $\omega\simeq u/l$. Using this, the frequency dependence of the diffusion coefficient can be translated into the corresponding $l$-dependence, the result being $D(l) \propto l^{-\eta}$. Balancing the rate of decay of the cluster with the outward flux of the particles we write $dl / dt \propto -l ^{-\eta}$. Integrating this simple equation over time from $t=0$ to $t=\tau_\lambda$ and over $l$ from $l=\lambda$ to $l=0$ one finds the dispersion relation $\tau_\lambda \propto \lambda ^{z}$ with $z = 1+\eta = 2-\gamma$. The persistency \cite{Feder} of relaxation is measured by the Hurst exponent $H$, which is related to our $z$ via $H = 1/z$. Lastly, the $\tau$-exponent, which defines the distribution of the particle flows caused by a single chain reaction, or activation-cluster size distribution \cite{Zhang}, is obtained from Eq.~(5) of Ref. \cite{Tang}, where one replaces $g$ with $\alpha$, and the fractal dimension $D$ with the fractal dimension of the infinite percolation cluster, $d_f = d - \beta / \nu$. The end result is $\tau = 3-\alpha z / d_f$. Note that the $\tau$ values in Refs. \cite{Tang} and \cite{Zhang} differ by 1. 

\subparagraph{Occurrence frequency energy distribution and the ${\ss}$-exponent} It is convenient to think of the activation clusters as containing a certain amount of ``energy" which is released when the comprising free particles dissipate to the boundaries. Using here that the electric charge of the free particles is a conserved quantity, we may associate the energy content of the activation clusters with their electric charge content. The latter $-$ accounting that each site of the activation cluster contains only one particle $-$ will be proportional to the fractal volume $\sim \lambda^{d_f}$, with $\lambda$ the size of the cluster. More so, the energy confinement time, $\tau_\lambda$, can be obtained as the activation cluster lifetime, enabling $\tau_\lambda \propto \lambda ^{z} \propto \epsilon_\lambda^{z/d_f}$, where the dispersion relation has been used. Differentiating in Eq.~(\ref{Skew2}), one arrives at the inverse-power distribution
\begin{equation}
N(\epsilon_\lambda) \propto dw_{\gamma} (\epsilon_\lambda) / d\epsilon_\lambda \propto \epsilon_\lambda^{-{\ss}}, \label{OFED} 
\end{equation}
where ${\ss}=1+\eta z / d_f$ is a power-law slope and $d_f = d - \beta / \nu$ in the ``hyperuniversal" fractal dimension. We interpret this distribution as the familiar from applications \cite{Char,Hudson} occurrence frequency energy distribution. Indeed the distribution in Eq.~(\ref{OFED}) might find its significance in the statistics of solar flares and gamma bursts, provided that the underlying physics processes refer to SOC. We estimate the ${\ss}$ values shortly.      

\subparagraph{Values of the critical exponents} Using known estimates \cite{Stauffer,Isi,Naka} of the percolation indices $\beta$, $\nu$, and $\mu$ we could evaluate the critical exponents of the DPRW model in all ambient dimensions $d\geq 1$. The results of this evaluation, summarized in Table~1, are in good agreement with the reported numerical values from the traditional sandpiles (for $d=2$, $z\approx 1.29$, $\tau\approx 2.0$; for $d=3$, $z\approx 1.7$, $\tau\approx 2.33$) \cite{Tang} and earlier theoretical predictions (for $d=2$, $z = 4/3$, $\tau = 2$; for $d=3$, $z = 5/3$, $\tau = 7/3$) \cite{Zhang}. We consider this conformity as a manifestation of the universality class of the model. For $d=\infty$, the model reproduces the exponents of mean-field SOC \cite{MF}.

In many ways, the DPRW approach to SOC offers a simple yet relevant lattice model for dielectric relaxation phenomena in systems with spatial disorder. One by-product of this approach is the case for stretched-exponential the KWW relaxation function, Eq.~(\ref{KWW}), which is often found empirically in various amorphous materials as for instance in many polymers and glass-like materials near the glass transition temperature (for reviews see Refs. \cite{Phillips,Kaatz}; references therein). In this connection, we observe that the model gives values of the exponent $\gamma$ (for $d=2$, $\gamma\approx 0.66$; for $d=3$, $\gamma\approx 0.4$) in good agreement with the typical experimental results (the $\gamma$ value between 0.3 and 0.8) \cite{Phillips,Montroll,Rolla,Jacobs}. This observation supports the hypothesis \cite{PRB2007,PLA2008} that dielectrics exhibiting stretched exponential relaxations are in a state of self-organized criticality. 

More so, the DPRW model gives a Hurst exponent (for $d=2$, $H\approx 0.75$; for $d=3$, $H\approx 0.6$) consistently with the reported narrow range of variation of $H$ as observed in different magnetic confinement systems (Hurst exponent varying between $H \approx 0.62$ and $0.75$) \cite{Carreras,Pedrosa,Carreras2,Carreras3}. In this connection, it is worth remarking that SOC behavior of the bulk plasma transport is expected to be a characteristic of higher-power plasma discharges in the so-called low confinement regime \cite{Carreras}.   

With respect to the occurrence frequency energy distribution, Eq.~(\ref{OFED}), the model predicts that ${\ss}=1$ in one dimension (for $d=1$, $\eta=0$) and ${\ss}=3/2$ in the mean-field limit (for $d\geq 6$, $z=2$, $\eta = 1$, and $d_f = 4$). These results are exact. Also, one finds, approximately, ${\ss}\approx 1.24$ for $d=2$ and ${\ss}\approx 1.4$ for $d=3$ (Hausdorff fractal dimensions $d_f=91/48$ and $d_f\approx 2.5$, accordingly). These theoretical predictions are in close agreement with the predictions of Aschwanden \cite{Ash12} who found values of ${\ss}=1.0$ for $d=1$; ${\ss}=1.28$ for $d=2$; and ${\ss}=1.5$ for $d=3$ from a statistical fractal-diffusive avalanche model of a slowly-driven self-organized criticality system (Ch. 2.2.2, this volume. In the notation of Ref. \cite{Ash12}, ${\ss} = \alpha_E$). The latter model assumes that the avalanche size grows as a diffusive random walk, implying that $z=2$ in all dimensions $d=1, 2, 3$. In our model, diffusive random walks are recovered in the mean-field limit only and in relatively high ambient dimensions that are not lesser than 6, while the $z$ exponent is taken to be non-diffusive in general. Even so, this does not seem to affect the ${\ss}$ value significantly, such that our mean-field result, ${\ss}=3/2$, almost precisely coincides with the result of Aschwanden \cite{Ash12} in three dimensions. Earlier, Litvinenko \cite{Lit} has suggested that the distribution of flare energies is characterized by a power-law with the slope ${\ss}=3/2$ independently of the ambient dimensionality $d > 1$. He modeled an avalanching process on a tree without loops, thus giving rise to this value. In this context, we should stress that the effect of loops can be abandoned only in the high dimensions $d\geq 6$, permitting a mean-field description \cite{Naka,Havlin}. All in all, the exponent ${\ss}=3/2$ agrees well with the reported slopes of the occurrence frequency energy distribution for solar flares (around $-1.5$ to $-1.8$) \cite{Hudson,Crosby,Uchida}, demonstrating that the observed power-law distribution of flare energy release is well reproduced under the assumption that the solar corona operates as a self-organized criticality system \cite{Char,Ash11,Ash12,Vlahos95,Norman}.

%
%
%
\begin{table}
\caption{Critical exponents of the DPRW model. Exponents appearing in statistical distributions as for instance inverse power-law power spectral density distribution are summarized in the lower part of the table. The mean-field results, holding for $d\geq 6$, are collected as $d=\infty$. Input parameters are the percolation indices $\beta$, $\nu$, and $\mu$ \cite{Isi,Stauffer,Naka}.}
\label{tab:1}       
%
%
\begin{tabular}{p{1.4cm}p{2.2cm}p{3.5cm}p{1cm}p{1cm}p{1cm}p{1cm}}
\hline\noalign{\smallskip}
Exponent & Expression & Description & $d=1$ & $d=2 $ & $d=3 $ & $d=\infty$ \\
\noalign{\smallskip}\svhline\noalign{\smallskip}
$\eta$& $\mu / (2\nu + \mu - \beta)$ & ac conductivity & $0$& $0.34$& $0.6$& $1$ \\
$z$& $1+\eta$ & Dispersion relation & $1$& $1.34$& $1.6$& $2$ \\
$\gamma$& $1-\eta$ & Mittag-Leffler relaxation & $1$& $0.66$& $0.4$& $0$ \\
$H$& $1/z$& Hurst exponent & $1$& $0.75$& $0.6$& $1/2$ \\
\hline\noalign{\smallskip}
Exponent & Expression & Distribution & $d=1$ & $d=2 $ & $d=3 $ & $d=\infty$ \\
\noalign{\smallskip}\svhline\noalign{\smallskip}
$\eta$& $\mu / (2\nu + \mu - \beta)$ & Relaxation-time & $0$& $0.34$& $0.6$& $1$ \\
$\alpha$& $2-2\eta$ & Power spectral density & $2$& $1.3$& $0.8$& $0$ \\
$\tau$& $3-\alpha z / d_f$ & Activation-cluster size & $1$& $2.1$& $2.5$& $3$ \\
${\ss}$& $1+\eta z / d_f$ & Occurrence frequency energy & $1$& $1.24$& $1.4$& $3/2$ \\
\noalign{\smallskip}\hline\noalign{\smallskip}
\end{tabular}
\end{table}

\section{Discussion} 

\subparagraph{General} Apart from details of the mathematical formalism, the DPRW model is actually quite simple. The main points are as follows. A lattice site can either be empty or occupied. An occupied site is interpreted as polarization charge. The equilibrium concentration of the polarization charges depends on the potential difference between the plates. When the potential difference changes the lattice occupancy parameter adjusts. A dynamical mechanism for this uses holes. The holes are just missing polarization charges. They are important key elements to the model as they provide a mechanism for the polarization current in the system. Beside holes, the free charges are introduced. The free charges, too, carry electric current whose very specific role in the model is just to control the potential difference between the plates. The changing amount of the free charges in the system has effect on the lattice occupancy parameter. Nonlinearly, it affects the conductivity of the lattice. This nonlinear twist provides a dynamical feedback by which the system is stabilized at the state of critical percolation. In many ways the proposed model is but a simple lattice model for dielectric relaxation in a self-adjusting disordered medium. It is perhaps the simplest model which accounts for the whole set of relaxation processes including the hole conduction. 

It is worth assessing the advantages and disadvantages of the DPRW approach to SOC. In terms of advantages, the electric nature of the model greatly facilitates the analytical theory: Not only does it permit to quantify the microscopic lattice rules in terms of the frequency-dependent complex ac conductivity, the use of the Kramers-Kronig relation in Eq.~(\ref{KK}) makes it possible to directly obtain the susceptibility function by integrating the conductivity response. As a result, the exponents $z$, $\gamma$, $\alpha$, and $H$ are expressible in terms of only one parameter, the exponent of ac conduction $\eta$. The latter is obtained as a simple function of the percolation indices $\beta$, $\nu$, and $\mu$.   

With respect to disadvantages, the model is seemingly different from the traditional approaches to SOC based on cellular automation (CA) and its integration in the existing family of SOC models might be a matter of debate. Even so, the idea of the random walks on a self-organized percolation system as a simplified yet relevant model for SOC constitutes a significant appeal: First, it relies on the established mathematical formalism of the random walks \cite{Havlin,Bouchaud,Klafter,Klafter2} whose advance on the SOC problem is theoretically very beneficial. Second, it offers a clear connection to studies outside the conventional SOC paradigm as for instance to transport of mass and charge in disordered media \cite{Dyre,Lax,Lax58,Druger}. Instead, the traditional CA type models are complicated by a poor analytical description of the microscopic transport mechanisms and their basic physics appreciation is at times uneasy.  

\subparagraph{The role of random walks} It is theoretically important to note that the dielectric context of the considered model, apart from offering a convenient platform for the analytical theory, is not however essential for the SOC phenomena. Indeed the DPRW model could be defined in terms of diffusion processes for neutral particles of different kinds. A formal reason for this is the equivalence \cite{Lax,Lax58} of the frequency-dependent electrical conductivity problem and the frequency-dependent diffusion problem, specific to hopping conduction. The crucial element to the model is the assumption of the random walks, not the nature of the particles. 

The possible generalizations of the DPRW model correspond to biased random walks of the free charges in the direction of the potential drop and/or inclusion of a second critical threshold $p_{cc} \geq p_c$ above which the random walk dynamics might change to a biased motion. We consider those generalizations obvious as they mainly intend to modify the value of the exponent $\eta$ in a certain parameter range, while the basic physical picture of SOC will remain essentially the same.    

\subparagraph{Universality class} The final point to be addressed here concerns the issue of {universality class}. We take notice of the fact that the DPRW SOC model uses the {\it charitable} redistribution rule \cite{Maslov} to propagate the activities, likewise to the traditional BTW sandpile \cite{BTW} or similar \cite{Tang,Zhang}. That means that an active site always loses its content to the neighbors. The charitable rule is to be distinguished from the {\it neutral} rule, when each of $2d + 1$ sites involved in redistribution gets an unbiased random share of the transported quantity. Models using the neutral rule often fall in the universality class of directed percolation (DP) and are characterized by appreciably larger values of the dynamic exponent $z$ (for $d=2$, $z\approx 1.73\pm0.05$) \cite{Maslov}. Based on this evidence, we suggest that the DPRW model belongs to the same universality class as the BTW sandpile, and not to the DP universality class, consistently with the values of the critical exponents collected in Table~1. 

\section{``Sakura" model} 

The DPRW model of SOC can be extended so that it includes the phenomena of self-organized ``turbulence" in Earth's geomagnetic tail, discussed in Refs. \cite{Japan,JGR96,JGR}. The main idea here is that, when the processes of magnetic reconnection stretch the magnetotail beyond a certain limit, the cross-tail electric current system of Earth's dipole-like magnetic field is destabilized, since it crucially requires the presence of a regular component of the geomagnetic field normal to the current sheet plane. Then the magnetotail spontaneously evolves into a far from equilibrium dynamical ``turbulent" state,  where it responds to changes in the tail current intensity and the varying dawn-dusk potential difference in terms of turbulent perturbation electric currents and magnetic field fluctuations. It was argued that the transport of electric charge across the magnetotail was due to heavier plasma species, the ions, whose regular orbits (transient or Speiser, weakly trapped, trapped) were essentially destroyed in the absence of stabilizing normal magnetic field and that the steady state which is self-organized corresponded to a strongly shaped electric current system and to a highly inhomogeneous, multi-scale magnetic fluctuation pattern (Fig.~5).  

This model for the coupled turbulent perturbation electric currents and magnetic field fluctuations has come to be known as ``Sakura" model, after its presentation \cite{Japan} at the Chapman Conference held in Kanazawa, Japan, November 5-9, 1996. Applications of the Sakura model pertain to the phenomena of tail current disruption in substorm regions of the near-Earth tail \cite{JGR,JASTP} as well as to the explanation of permanent presence of magnetic field fluctuations in the distant magnetotail \cite{UFN,JGR96}, as suggested by the GEOTAIL measurements \cite{Hoshino,Nishida}. The important role of the ion component of the plasma in driving cross-tail current instability was addressed in Refs. \cite{Ohtani,Lui,Malova}, where one also finds an analysis of the satellite observational data.

\begin{figure}
\includegraphics[width=1.10\textwidth]{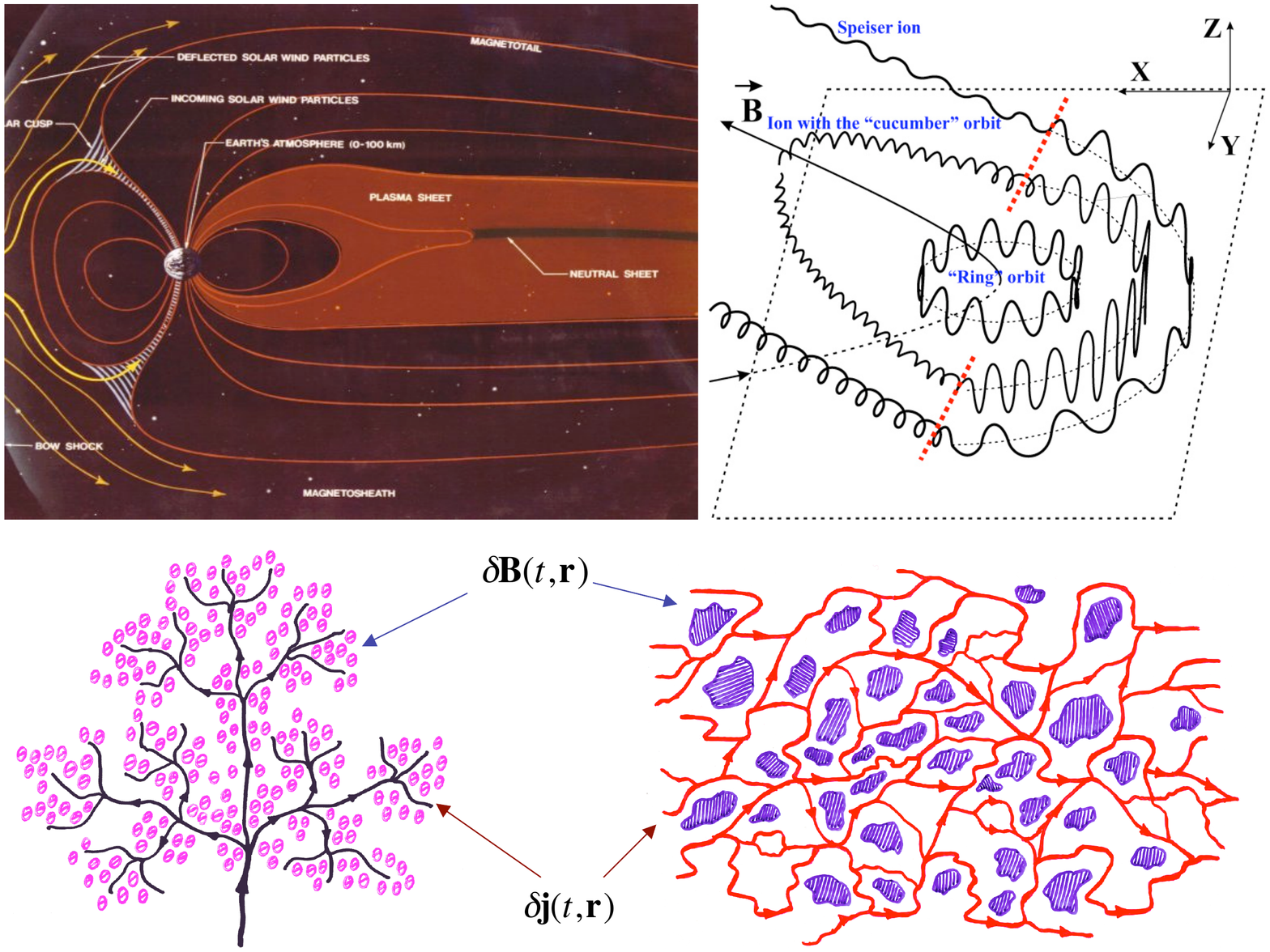}
\caption{\label{} ``Sakura" model. Top left: Schematic illustration for the Earth's dynamic magnetosphere interacting with its solar wind drive. Top right: The various orbit types for thermal ions in the magnetotail current sheet with finite regular component of the normal field. Adapted from Ref. \cite{Malova}. When the processes of magnetic reconnection stretch the magnetotail beyond a certain limit, the particle regular orbital motion is essentially destabilized and the cross-tail electric current system spontaneously evolves into a far from equilibrium dynamical ``turbulent" state where it responds to changes in the tail current intensity and the varying dawn-dusk potential difference in terms of turbulent perturbation electric currents and magnetic field fluctuations. Bottom left: Original model promotion as a blossoming sakura tree: Its ``leaves" represent the magnetic field fluctuations; its ``branches," the perturbation electric currents. Bottom right: Plan view of the turbulent current sheet. Cross-tail electric currents are organized in highly branched, very inhomogeneous conducting patterns with the topology of a fractal network at percolation (red color). The magnetic field fluctuations are shown as chunks of different sizes (violet color). They scatter the momentum of current-carrying particles, the ions, and are electromagnetically related with the perturbation electric current intensity by means of Maxwell's equation, ${\bf \nabla}\times\delta {\bf B} (t, {\bf r}) = (4\pi / c) \delta {\bf j} (t, {\bf {r}})$. Note that the magnetic field fluctuations, $\delta {\bf B} (t, {\bf r})$, can be thought as analog polarization charges in the DPRW model of SOC.}
\end{figure}

Denoting the perturbation tail current density and magnetic fluctuation field by respectively $\delta {\bf j} (t,{\bf r})$ and $\delta {\bf B} (t,{\bf r})$, we write  
\begin{equation}
{\bf \nabla}\times\delta {\bf B} (t,{\bf r}) = \frac{4\pi}{c} \delta {\bf j} (t, {\bf {r}}), \label{Sakura} 
\end{equation}
where, under the assumptions of locality and linear conductivity response,
\begin{equation}
\delta {\bf j} (t,{\bf r}) = \int_{-\infty}^{+\infty} \sigma (t - t^{\prime}) \delta {\bf E} (t^{\prime}, {\bf {r}}) dt^{\prime}. \label{CDF} 
\end{equation} 
Here, $\delta {\bf E} (t, {\bf {r}})$ is the perturbation electric field in the current sheet plane. The memory function, $\sigma (t)$, is obtained as Fourier inversion of the frequency-dependent complex ac conductivity, $\sigma_{\rm ac} (\omega)$, where the ac dependence is due to the turbulent nature of the conducting domain. It is understood that the time varying magnetic perturbation generates a time and spatially varying electric field because of Faraday's law. An important feature which arises in this induction process is gradual heating and energization of the plasma, leading to the occurrence of slowly decaying, high-energy non-thermal wings in the particle energy distribution function \cite{UFN,PRE01+,Adv2002}. Often in the magnetospheric plasma research those distributions with wings are modeled by non-thermal the so-called ``kappa" distributions \cite{Christon} which interpolate between the initial low-energy exponential forms and the asymptotic inverse power-law behavior. The significance of the ``kappa" distributions lies in the fact \cite{NPG} that they appear as canonical distributions in the non-extensive thermodynamics due to Tsallis (Ch. 2.3.5, this volume) \cite{Tsallis}. In the present analysis we shall assume, however, that all fluctuation frequencies are so slow that the effect of the inductive field can be neglected and we omit, consequently, the inductive term in Eq.~(\ref{Sakura}) above. To this end, performing a Fourier transform of Eq.~(\ref{Sakura}) in space and time, we have
\begin{equation}
{\bf k}\times\delta {\bf B} (\omega, {\bf k}) = \frac{4\pi}{c} \delta {\bf j} (\omega, {\bf {k}}), \label{FSakura} 
\end{equation}
where ${\bf k}$ is the wave vector of the perturbation. Simultaneously, from Eq.~(\ref{CDF}) we find $\delta {\bf j} (\omega, {\bf k}) = \sigma_{\rm ac} (\omega) \delta {\bf E} (\omega, {\bf k})$. In vicinity of the current instability threshold, considering the low-frequency limit of Eq.~(\ref{Sakura}), we may think of the fluctuations as of collection of plane waves, characterized by a linear dispersion relation $\omega = {\bf k}\cdot {\bf u}$, where the phase velocity, ${\bf u}$, does not depend on ${\bf k}$. Given that the input perturbing electric field is uncorrelated white noise: $\delta {\bf E} (\omega, {\bf {k}}) = \bf 1$, with the aid of Eq.~(\ref{FSakura}) one sees that the power spectral density of the magnetic fluctuation field, $|\delta {\bf B} (\omega, {\bf k})|^2$, will be proportional to $S(\omega) \propto |\sigma_{\rm ac} (\omega) / \omega|^2$. At this point, if one assumes, following Refs. \cite{Japan,JGR}, that the dynamical ``turbulent" state is characterized by fractal geometry of the threshold percolation, one may exploit the scaling relation $\sigma_{\rm ac} (\omega) \propto \omega^\eta$ to obtain $S (\omega) \propto \omega^{-\alpha}$, with $\alpha = 2(1-\eta)$, consistently with the DPRW result. Utilizing the percolation estimate above (for $d=2$, $\eta\approx 0.34$), one finds that $\alpha\approx 1.3$. This theoretical prediction compares well against the reported $\alpha$ values in the lower-frequency part of the magnetic fluctuation spectrum (below a turnover or knee frequency posed by the unstable tearing modes: see Fig.~6) \cite{Hoshino,Ohtani,Bauer}. Thus, behavior is self-similar in the self-organization domain. Remark that the power spectral density of the magnetic fluctuation field, $|\delta {\bf B} (\omega, {\bf k})|^2$, corresponds with the power spectral density of the dynamic polarization response, Eq.~(\ref{PSD}). We should stress that the Sakura model leads to a smaller spectral index ($\alpha\approx 1.3$) than Kolmogorov's theoretical value for fluid turbulence ($\alpha = 5/3$) as well as Kraichnan's theoretical value for magnetohydrodynamic plasma turbulence ($\alpha = 3/2$). This last observation addresses the significance of ``self-organized" magnetic fluctuation turbulence as opposed to ordinary (fluid-like) turbulence.

\begin{figure}
\includegraphics[width=1.0\textwidth]{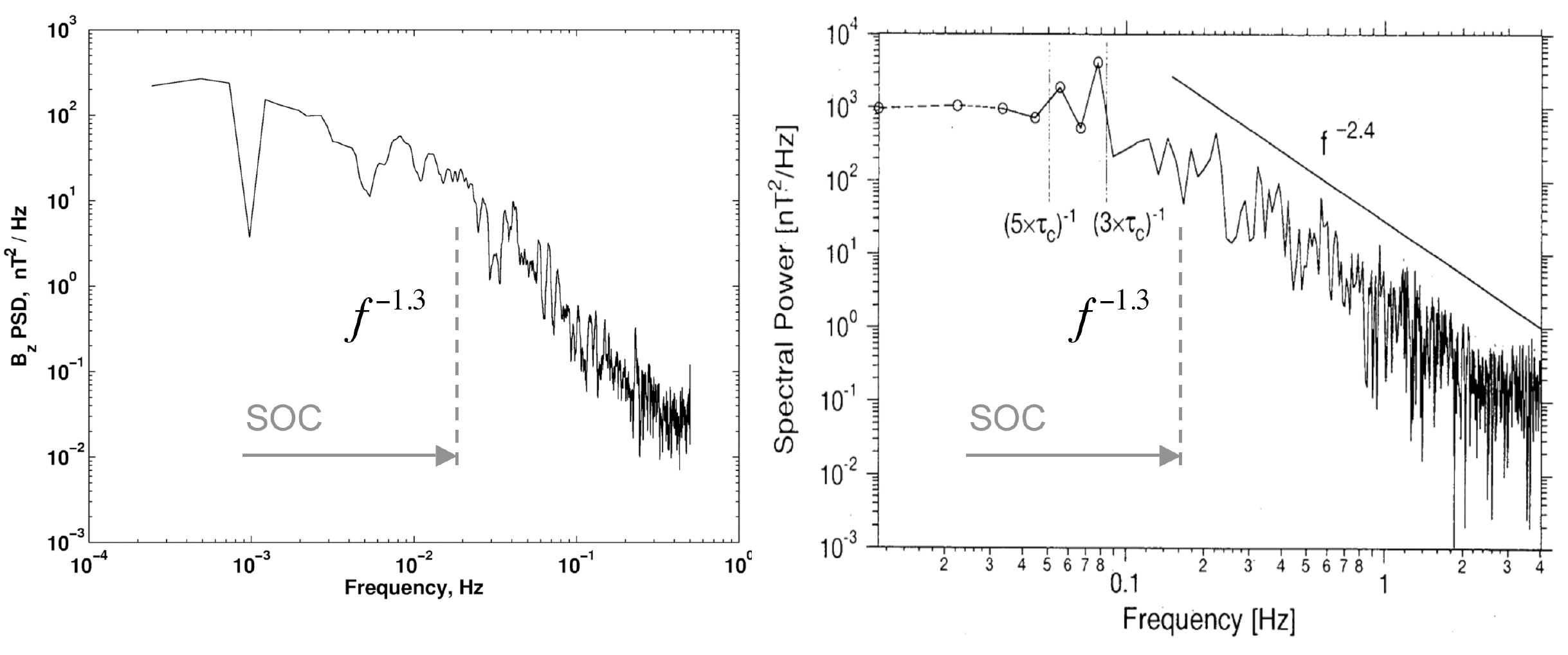}
\caption{\label{} The typical spectra of magnetic field fluctuations in the near-Earth stretched and thinned magnetotail prior to tail current disruption. Left: INTERBALL-1 observations. Courtesy of prof. Lev Zelenyi. Right: A spectrum observed by the Charge Composition Explorer of the Active Magnetospheric Particle Tracer Explorers (AMPTE) satellite, with an emphasis on the August 28, 1986, current disruption event. Adapted from Ref. \cite{Ohtani}. A distinctive feature of these spectra is the knee around a characteristic frequency posed by the unstable tearing modes. The spectrum is flatter below the knee frequency and is noticeably steeper just above it. The processes of self-organization to a critical state correspond to the flatter counterpart of the spectrum and to frequencies lesser than the knee frequency: in practice, lesser than $\sim 5\cdot 10^{-2}$ Hz, although the exact bound, in real data, may not be that certain. Associated spectrum is a power-law with the typical slope $\sim -1.3$ (in log-log plot). In the limit of very low frequencies, the observed signals cross over to a white noise, as they should (this is due to finite system size effects and/or the natural limitations of the observational time series), so that the spectra are flat. The spectral properties of the fluctuations just above the knee (slope $\sim -2.4$) involve the processes of convection of magnetic turbulence structures with decelerated solar wind velocity along the magnetotail. These processes were analyzed in Refs. \cite{JGR,JASTP} and are not considered here. All in all, the signatures of self-organization to a critical state are to be expected in the intermediate frequency range, comprised between those frequencies where the spectra are white noise-like and the knee frequency. An approximate position of the knee is marked by vertical dashed line.}
\end{figure}

Our conclusion so far is that the power spectral density of the magnetic fluctuation field is given by an inverse power-law and that the behavior corresponds with the prediction of the DPRW SOC model. This result suggests that micro-processes of self-organization of electric currents and magnetic field fluctuations in the Earth's stretched and thinned magnetotail are governed by SOC \cite{UFN,JASTP}. Remark that the above assumption of fractality and self-similar behavior is validated through the direct analysis of time series of the satellite observed magnetic field fluctuations, obtained {\it in situ} in the substorm regions of the near-Earth tail and associated with the phenomena of tail current instability during the substorm growth phase \cite{Ohtani}.

\section{Beyond the linear theories: DANSE formalism}

Theoretical approaches discussed so far were based on a linear-response theory and on fractional generalizations of the diffusion and relaxation equations, Eqs.~(\ref{FRE}) and Eq.~(\ref{End}), which are linear by construction. Nonlinearities were contained in fractal geometry of the dynamical system at percolation and implicitly in the various ``avalanche" exponents and the fractional indices of time differentiation. In the present analysis, this paradigm of ``linear dynamics in a nonlinear medium" will be relaxed. Rather, a more general theoretical picture will be drawn, in which the dynamical and structural nonlinearities are twisted, and for which one might propose the formula ``nonlinear dynamics in a nonlinear medium." Very specifically, we intend to demonstrate that the phenomena of SOC $-$ at least those belonging to the universality class of the DPRW model $-$ may be cast in the mathematical formalism of discrete Anderson nonlinear Schr\"odinger equation (DANSE), which invokes a random potential for lattice interactions, and in which the strength of nonlinearity, being an inherent part of the model description, is determined {dynamically} as the system self-adjusts and evolves to criticality in response to external forcing. Most previous theoretical studies of SOC have neglected the possibility of describing the lattice interactions in terms of a random potential field, and have focused, consequently, on the microscopic redistribution rules for the dynamics. We believe that this approach is unnecessarily restrictive and has left out the important physics results. 

\subparagraph{The roadmap} As is already mentioned above, in the DPRW model the critical state is made self-organized via the mechanisms of hole-hopping by which the system responds to the fluctuating potential difference on the capacitor. We should stress that the holes redistribute the polarization charges in a way as to preserve the properties of the random percolation. They change the shape and the folding of the percolation clusters in the ambient configuration space, but not the random character in the distribution of the conducting sites. A confirmation of these ideas can be obtained if one considers holes as ``excitations" of self-organized critical state mediating the lattice activities. With this interpretation in mind, the transport problem for holes can be formulated as a transport problem for the hole wave function in a random potential field.\footnote{Essentially the same approach applies to the electrons.} The latter problem, in its turn, can be studied as part of the general problem of transport of waves in disordered media, with that very specific element that the medium is nonlinear and its properties are coupled with the wave process itself $-$ which, too, can be nonlinear. 

We consider the problem of dynamical localization of the hole wave function in the framework of nonlinear Schr\"odinger equation (NLSE) with a random potential term on a lattice. The NLSE was derived for a variety of physical systems under some approximations \cite{Segur}. Recently, it has been rigorously established that, for a large variety of physical conditions and details of interaction, the NLSE (also known as the Gross-Pitaevsky equation) is exact in the thermodynamic limit \cite{GPE}. An important feature which arises in this approach is competition between randomness and nonlinearity. As we shall see, this competition has a significant effect on SOC problem. Under most general conditions, we can expect that, when the nonlinearity is sufficiently small, the random properties dominate, giving rise to the phenomena of Anderson localization of the excitations. That means that diffusion is suppressed and a wave packet that is initially localized will not spread to infinity \cite{And}. 

So, what happens with the increasing strength of nonlinearity? The question is far than trivial, after the simulation-motivated suggestion \cite{Sh93,PS} that a weak nonlinearity can destroy localization above some level, giving rise to unlimited spreading of the wave field along the lattice, despite the existing disorder. The dynamics of the spreading has remained a matter of debate \cite{PS,Flach,WangZhang,Iomin,Fishman}. 

In a recent investigation of NLSE with disorder, it has been theoretically found \cite{EPLI} that destruction of Anderson localization in the presence of nonlinearity is a critical phenomenon $-$ similar to a percolation transition in random lattices. Delocalization occurs spontaneously when the strength of nonlinearity goes above a certain limit. Below that limit, the field is localized similarly to the linear case. In the analysis of this section, we bring these ideas in contact with the physics of SOC and we consider a situation in which the strength of nonlinearity is determined {dynamically} in terms of time depending probability of site occupancy as the system fluctuates near the critical point. It is this very specific nonlinear twist with the dynamical state of the lattice, which captures the essential key signatures due to SOC, and which as is suggested below enables to cast the SOC problem into the mathematical formalism of a discrete NLSE with disorder. The effect this twist has on the dynamics is that the localization-delocalization transition is turned self-organized, occurring exactly at the state of critical percolation. 

With the aid of our DPRW lattice model, we can essentially simplify the analysis if we consider that the transport in vicinity of the critical state occurs as a result of hole hopping between the nearest-neighbor sites, occupied by the polarization charges. As the percolation threshold is approached, we can envisage clusters of the polarization charges as the effective (random) medium, acting as a random potential on the hole wave function. Following Anderson \cite{And}, we adopt the tight binding description for the hopping processes. Consequently, we introduce a Hamiltonian, paving the way to consider the transport problem for SOC as essentially a Hamiltonian problem. This approach poses a theoretical challenge, as it aims to connect SOC with first-principle models.  

\subparagraph{DANSE equation} Focusing on the hopping motions on a lattice, we consider a variant of discrete Anderson nonlinear Schr\"odinger equation (DANSE) with randomness
\begin{equation}
i\hbar\frac{\partial\psi_n}{\partial t} = \hat{H}_L\psi_n + \zeta |\psi_n|^2 \psi_n,
\label{DANSE} 
\end{equation}
where 
\begin{equation}
\hat{H}_L\psi_n = \varepsilon_n\psi_n + V (\psi_{n+1} + \psi_{n-1}),
\label{TB} 
\end{equation}
$\hat H_L$ is the Hamiltonian of the linear problem in the tight-binding approximation \cite{And}; $\psi_n$ is the hole wave function; $\hat{H}_L\psi_n$ describes hopping-like transitions between the nearest-neighbor sites on a lattice; and $\zeta |\psi_n|^2 \psi_n$ accounts for the generic nonlinearity of the wave process. In the above, $\zeta$ characterizes the strength of nonlinearity;\footnote{We assume that the nonlinearity is repulsive ($\zeta > 0$), implying that it favors the spreading of the wave field. In the opposite case of attractive nonlinearity ($\zeta < 0$), solitons are typically found \cite{UFN,Segur}.} on-site energies $\varepsilon_n$ are randomly distributed with zero mean across a finite energy range; $V$ is hopping matrix element; and the total probability is normalized to $\sum_n |\psi_n|^2 = 1$. More general models can be obtained by replacing $|\psi_n|^2$ with $|\psi_n|^r$, for arbitrary $r > 0$. We do not consider such models here. In what follows, $\hbar = 1$ for simplicity. When $\zeta\rightarrow 0$, all eigenstates are exponentially localized \cite{And}. In the absence of randomness, DANSE~(\ref{DANSE}) is completely integrable. Adhering to the effective-medium description, we aim to comprehend the spreading of the hole wave function under the action of nonlinear term in the limit $t\rightarrow +\infty$.   

\subparagraph{Coupled nonlinear oscillators} It is useful to expand the hole wave function using an orthogonal basis of the eigenstates, $\phi_{n,m}$, of the Anderson Hamiltonian, $\hat H_L$, yielding
\begin{equation}
\psi_n = \sum_m \sigma_m (t) \phi_{n,m} \ \ \ (m= 1,2,\dots). 
\label{3ap} 
\end{equation}
``Orthogonal" means that $\sum _n \phi^*_{n,m}\phi_{n,k} = \delta_{m,k}$, where $\delta_{m,k}$ is Kronecker's delta and star denotes complex conjugate. The total probability being equal to 1 implies that $\sum_n \psi_n^*\psi_n = \sum_m \sigma_m^* (t)\sigma_m (t) = 1$. For the nonlinear equation, Eq.~(\ref{DANSE}), the dependence of the expansion coefficients, $\sigma_m (t)$, is found to be\footnote{Hint: substitute Eq.~(\ref{3ap}) into DANSE~(\ref{DANSE}), then multiply the both sides by $\phi^*_{n,k}$, and sum over $n$, remembering that the modes are orthogonal.} 
\begin{equation}
i\dot{\sigma}_k - \omega_k \sigma_k = \zeta \sum_{m_1, m_2, m_3} V_{k, m_1, m_2, m_3} \sigma_{m_1} \sigma^*_{m_2} \sigma_{m_3},
\label{4ap} 
\end{equation}
where dot denotes time differentiation; $\omega_k$ $(k= 1,2,\dots)$ are the eigenfrequencies of $\hat H_L$; and the amplitudes $V_{k, m_1, m_2, m_3}$ are given by
\begin{equation}
V_{k, m_1, m_2, m_3} = \sum_{n} \phi^*_{n,k}\phi_{n,m_1}\phi^*_{n,m_2}\phi_{n,m_3}.
\label{5ap} 
\end{equation}
In deriving Eq.~(\ref{4ap}) we took into account that $\hat H_L \phi_{n,k} = \omega_k \phi_{n,k}$. Equations~(\ref{4ap}) correspond to a system of coupled nonlinear oscillators with the Hamiltonian 
\begin{equation}
\hat H = \hat H_{0} + \hat H_{\rm int}, \ \ \ \hat H_0 = \sum_k \omega_k \sigma^*_k \sigma_k,
\label{6ap} 
\end{equation}
\begin{equation}
\hat H_{\rm int} = \frac{\zeta}{2} \sum_{k, m_1, m_2, m_3} V_{k, m_1, m_2, m_3} \sigma^*_k \sigma_{m_1} \sigma^*_{m_2} \sigma_{m_3}.
\label{6+ap} 
\end{equation}
Thus we have translated the hopping problem for the hole wave function into the interaction problem for coupled nonlinear oscillators on a lattice. In the above, $\hat H_{0}$ is the Hamiltonian of non-interacting harmonic oscillators and $\hat H_{\rm int}$ is the interaction Hamiltonian.\footnote{We include self-interactions into $\hat H_{\rm int}$.} Each nonlinear oscillator with the Hamiltonian   
\begin{equation}
\hat h_{k} = \omega_k \sigma^*_k \sigma_k + \frac{\zeta}{2} V_{k, k, k, k} \sigma^*_k \sigma_{k} \sigma^*_{k} \sigma_{k}
\label{6+hap} 
\end{equation}
and the equation of motion 
\begin{equation}
i\dot{\sigma}_k - \omega_k \sigma_k - \zeta V_{k, k, k, k} \sigma_{k} \sigma^*_{k} \sigma_{k} = 0
\label{eq} 
\end{equation}
represents one nonlinear eigenstate in the system $-$ identified by its wave number $k$, unperturbed frequency $\omega_k$, and nonlinear frequency shift $\Delta \omega_{k} = \zeta V_{k, k, k, k} \sigma_{k} \sigma^*_{k}$. Non-diagonal elements $V_{k, m_1, m_2, m_3}$ characterize couplings between each four eigenstates with wave numbers $k$, $m_1$, $m_2$, and $m_3$. It is understood that the excitation of each eigenstate is not other than the spreading of the wave field in wave number space. Resonances occur between the eigenfrequencies $\omega_k$ and the frequencies posed by the nonlinear interaction terms. We have\footnote{Conditions for nonlinear resonance are obtained by accounting for the nonlinear frequency shift.}
\begin{equation}
\omega_k = \omega_{m_1} - \omega_{m_2} + \omega_{m_3}.
\label{Res} 
\end{equation}
When the resonances happen to overlap, a phase trajectory may occasionally switch from one resonance to another. As Chirikov realized \cite{Chirikov}, any overlap of resonances will introduce a random element to the dynamics along with some transport in phase space. Applying this argument to DANSE~(\ref{DANSE}), one sees that destruction of Anderson localization is limited to a set of resonances in a Hamiltonian system of coupled nonlinear oscillators, Eqs.~(\ref{6ap}) and~(\ref{6+ap}), permitting a connected escape path to infinity. 

\subparagraph{Chaotic vs pseudochaotic dynamics} At this point, the focus is on topology of the random motions in phase space. We address an idealized situation first where the overlapping resonances densely fill the space. This is the familiar fully developed chaos, a regime that has been widely studied and discussed in the literature (e.g., Refs. \cite{ZaslavskyUFN,Sagdeev}). A concern raised over this regime when applied to Eqs.~(\ref{6ap}) and~(\ref{6+ap}) comes from the fact that it requires a diverging free energy reservoir in systems with a large number of interacting degrees of freedom such as SOC systems. Yet, developed chaos offers a simple toy-model for the transport as it corresponds with a well-understood, diffusive behavior. 

A more general, as well as more intricate, situation occurs when the random motions coexist along with regular (KAM regime) dynamics. If one takes this idea to its extreme, one ends up with the general problem of transport along separatrices of dynamical systems \cite{Arnold}. This problem constitutes a fascinating nonlinear problem that has as much appeal to the mathematician as to the physicist. An original important promotion of this problem to ``large" (spatially extended) systems is due to Chirikov and Vecheslavov \cite{ChV}.

This type of problem occurs for slow frequencies. One finds \cite{PRE01,PRE09} that resonance-overlap conditions are satisfied along the ``percolating" orbits or separatrices of the random potential where the orbital periods diverge. The available phase space for the random dynamics can be very ``narrow" in that case. In large systems, the set of separatrices can moreover be geometrically very complex and strongly shaped. Often it can be envisaged as a fractal network at percolation as for instance in random fields with sign-symmetry \cite{Efros,Isi,PRE00}. 

There is a fundamental difference between the above two transport regimes (chaotic vs near-separatrix). The former regime is associated with an exponential loss of correlation permitting a Fokker-Planck description in the limit $t\rightarrow+\infty$. The latter regime when considered for large systems is associated with an algebraic loss of correlation instead \cite{Report,Chaos,PhysicaD}, implying that the correlation time is infinite. There is no a conventional Fokker-Planck equation here, unless generalized to fractional derivatives \cite{Ctrw1,Ctrw2}, nor the familiar Markovian property (i.e., that the dynamics are memoryless). On the contrary, there is an interesting interplay between randomness, fractality, and correlation, which is manifest in the fact that all Lyapunov exponents vanish in the thermodynamic limit, despite that the dynamics are intrinsically random \cite{PRE09}. 

This situation of random non-chaotic dynamics with zero Lyapunov exponents, being in fact very general, has come to be known as ``pseudochaos" \cite{Report,JMPB,PD2004}. One may think of pseudochaos as occurring ``at the edge" of stochasticity and chaos, thus separating fully developed chaos from domains with regular motions. There is a growing belief that the concept of pseudochaos offers the natural mathematical platform to obtain the fractional kinetic equations from first principles \cite{Report}. 

In section 3 above it was argued that the phenomena of SOC could be described by fractional kinetics, which is a suitable and powerful formalism for long-range correlated behavior. Here, we lay more stress on this argument by proposing that SOC processes are as a matter of fact {\it pseudochaotic}. In this way of thinking one naturally bridges the concepts of fractional kinetics, non-Markovian transport, and SOC. Then the inherent ``edge" character of pseudochaotic dynamics can be related with what one believes is the threshold nature of SOC processes \cite{BTW}. Support to this suggestion can be found in the results of Refs. \cite{EPL,NJP,PRE09}. 

\subparagraph{Nearest-neighbor rule} This idea of ``edge" behavior brings us to a model \cite{EPLI} where each nonlinear oscillator, Eq.~(\ref{6+hap}), can only communicate with the rest of the wave field via a nearest-neighbor rule. Indeed this is the marginal regime yet permitting an escape path to infinity. We associate this regime with the onset of delocalization. Clearly, the number of coupling links is minimized in that case. Note that the nearest-neighbor rule guarantees that all interactions are local, this being an essential key element to SOC. When summing on the right-hand-side, the only combinations to be kept are, for the reasons of symmetry, $\sigma_{k} \sigma^*_{k} \sigma_{k}$ and $\sigma_{k-1} \sigma^*_{k} \sigma_{k+1}$. We have
\begin{equation}
i\dot{\sigma}_k - \omega_k \sigma_k = \zeta V_{k} \sigma_{k} \sigma^*_{k} \sigma_{k} + 2\zeta V_k^\pm \sigma_{k-1} \sigma^*_{k} \sigma_{k+1},
\label{9ap} 
\end{equation} 
where we have also denoted for simplicity $V_k = V_{k, k, k, k}$ and $V_k^\pm = V_{k, k-1, k, k+1}$. Equations~(\ref{9ap}) define an infinite ($k= 1,2,\dots$) chain of coupled nonlinear oscillators where all couplings are local (nearest-neighbor-like). The interaction Hamiltonian in Eq.~(\ref{6+ap}) is simplified to   
\begin{equation}
\hat H_{\rm int} = \frac{\zeta}{2} \sum_{k} V_{k} \sigma^*_k \sigma_{k} \sigma^*_{k} \sigma_{k} + {\zeta}\sum_{k} V_k^\pm \sigma^*_k \sigma_{k-1} \sigma^*_{k} \sigma_{k+1}.
\label{6++ap} 
\end{equation}

\subparagraph{Pseudochaotic dynamics on a Cayley tree} We are now in position to introduce a simple lattice model for the transport. The key step is to observe that Eqs.~(\ref{9ap}) can be mapped on a Cayley tree where each node is connected to $c=3$ neighbors (here, $c$ is the coordination number). The mapping is defined as follows. A node with coordinate $k$ represents a nonlinear eigenstate, or nonlinear oscillator with the equation of motion~(\ref{eq}). There are exactly $c=3$ branches at each node: one that we consider ingoing represents the complex amplitude $\sigma^*_{k}$, and the other two, the outgoing branches, represent the complex amplitudes $\sigma_{k-1}$ and $\sigma_{k+1}$ respectively. These settings are schematically illustrated in Fig.~7.  

\begin{figure}
\includegraphics[width=0.89\textwidth]{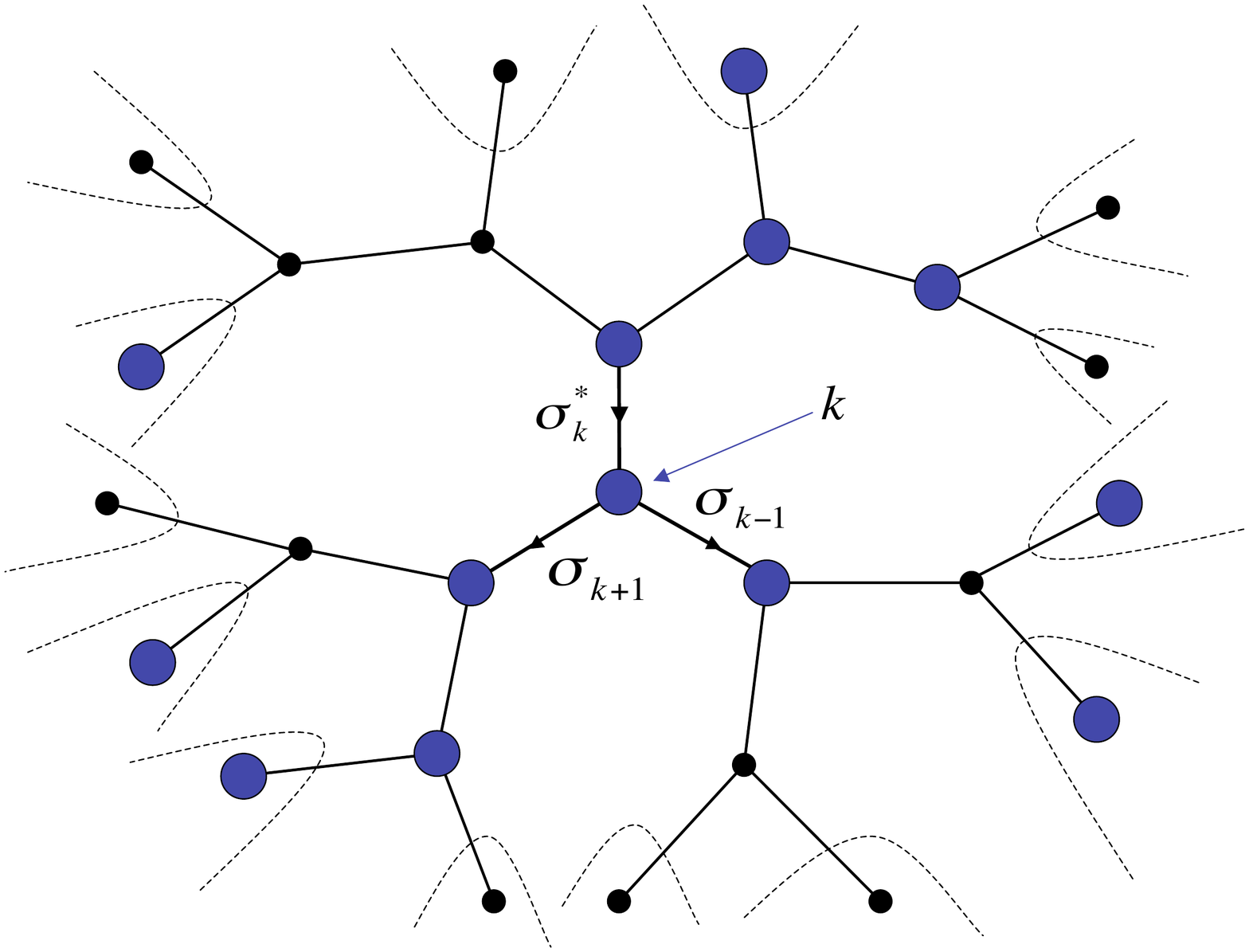}
\caption{\label{} Mapping Eqs.~(\ref{9ap}) on a Cayley tree. Each node represents a nonlinear eigenstate, or nonlinear oscillator with the equation of motion $i\dot{\sigma}_k - \omega_k \sigma_k - \beta V_{k, k, k, k} \sigma_{k} \sigma^*_{k} \sigma_{k} = 0$. Blue nodes represent oscillators in a chaotic (``dephased") state. Black nodes represent oscillators in regular state. One ingoing and two outgoing branches on node $k$ ($k= 1,2,\dots$) represent respectively the complex amplitudes $\sigma^*_{k}$, $\sigma_{k-1}$, and $\sigma_{k+1}$. Structures that are not explicitly shown are beyond the dashed lines. Adapted from Ref. \cite{EPLI}.}
\end{figure}

A Cayley tree being by its definition \cite{Schroeder} a hierarchical graph offers a suitable geometric model for infinite-dimensional spaces. We think of this graph as embedded into phase space of the Hamiltonian system of coupled nonlinear oscillators, Eqs.~(\ref{6ap}) and~(\ref{6+ap}). In the thermodynamic limit $k_{\max}\rightarrow\infty$, in place of a Cayley tree, one uses the notion of a Bethe lattice.\footnote{A Bethe lattice is an infinite version of the Cayley tree. To this end, a purist might prefer to say ``bond" in place of ``branch," but that's all about the terminology.} Setting $k_{\max}\rightarrow\infty$, we suppose that each node of the Bethe lattice hosts a nonlinear oscillator, Eq.~(\ref{eq}). The bonds of the lattice, in their turn, can conduct oscillatory processes to their neighbors as a result of the interactions present. 

Next, we assume that each oscillator can be in a chaotic (``dephased") state with the probability $p$ (and hence, in a regular state with the probability $1-p$). The $p$ value being smaller than 1 implies that the domains of random motions occupy only a fraction of the lattice nodes. Whether an oscillator is dephased is decided by Chirikov's resonance-overlap condition $-$ which may or may not be matched on node $k$. We believe \cite{ChV} that in systems with many coupled degrees of freedom each such ``decision" is essentially a matter of the probability. The choice is random. Focusing on the $p$ value, we consider system-average nonlinear frequency shift 
\begin{equation}
\Delta\omega_{\,\rm NL} = \zeta\langle|\psi_n|^2\rangle_{\Delta n}
\label{Shift} 
\end{equation}
as an effective ``temperature" of nonlinear interaction. It is this ``temperature" that rules over the excitation of the various resonant ``levels" in the system. With this interpretation in mind, one writes $p$ as the Boltzmann factor 
\begin{equation}
p = \exp (-\delta\omega / \Delta\omega_{\rm NL}),
\label{B} 
\end{equation}
where $\delta\omega$ is the characteristic energy gap between the resonances. Expanding $\psi_n$ over the basis of linearly localized modes, it is found that 
\begin{equation}
\langle|\psi_n|^2\rangle_{\Delta n} = \frac{1}{\Delta n} \sum_n \sum_{m_1,m_2} \phi^*_{n,m_1}\phi_{n,m_2} \sigma^*_{m_1} \sigma_{m_2}.
\label{Mean} 
\end{equation}  
The summation here is performed with the use of orthogonality of the basis modes. Combining with Eq.~(\ref{Shift}),
\begin{equation}
\Delta\omega_{\,\rm NL} =  \frac{\zeta}{\Delta n}\sum_m \sigma^*_{m} \sigma_{m}.
\label{Upon} 
\end{equation}
The sum over $m$ is easily seen to be equal to 1 due to the conservation of the probability. Thus, $\Delta\omega_{\rm NL} = \zeta / \Delta n$. When the field is spread over $\Delta n$ states, the distance between the resonant frequencies behaves as $\delta \omega \sim 1/\Delta n$. We normalize units in Eq.~(\ref{DANSE}) to have $\delta \omega = 1/\Delta n$ exactly. One sees that 
\begin{equation}
p = \exp (-1/\zeta).
\label{NP} 
\end{equation}
For the vanishing $\zeta\rightarrow 0$, the Boltzmann factor $p\rightarrow 0$, implying that all oscillators are in regular state. In the opposite regime of $\zeta\rightarrow\infty$, $p\rightarrow 1$. That means that all oscillators are dephased and that the random motions span the entire lattice. 

There is a critical concentration, $p_c$, of dephased oscillators permitting an escape path to infinity for the first time. This critical concentration is not other than the percolation threshold on a Cayley tree. In a basic theory of percolation it is found that $p_c = 1/(c-1)$ (e.g., Ref.~\cite{Schroeder}). This is an exact result. For $c=3$, $p_c = 1/2$. We associate the critical value $p_c = 1/2$ with the onset of transport in the DANSE model, Eq.~(\ref{DANSE}). When translated into the $\zeta$ values the threshold condition reads 
\begin{equation}
\zeta_c = 1/\ln (c-1).
\label{Betac} 
\end{equation}
Setting $c=3$, we have $\zeta_c = 1/\ln 2 \approx 1.4427$. This value defines the critical strength of nonlinearity that destroys the Anderson localization. For the $\zeta$ values smaller than this, the localization persists, despite that the problem is nonlinear. When $\zeta \geq 1/\ln 2$, the localization is lost and the wave field spreads to infinity. 

Our conclusion so far is that destruction of Anderson localization is a thresholded phenomenon, which can be described as a percolation transition in a system of dephased oscillators on a Cayley tree (Bethe lattice). Delocalization occurs when the strength of nonlinearity, mathematically related with the concentration of dephased oscillators, Eq.~(\ref{NP}), exceeds a certain critical level. The critical point is exactly at $\zeta_c = 1/\ln 2$. 

\subparagraph{Making delocalization transition self-organized} With the recognition that, according to Eq.~(\ref{NP}), the nonlinearity parameter, $\zeta$, is twisted with the probability of site occupancy, $p$, the formalism of discrete Anderson nonlinear Schr\"odinger equation, Eq.~(\ref{DANSE}), allows for a representation, which makes it possible to include the phenomena of SOC. The main idea here is to think of $\zeta$ as a fluctuating parameter, which is defined dynamically as the system evolves to percolation, and whose value is decided ``on-the-fly" by the actual state of the lattice. Then the $\zeta$ value need not fine tuned to its critical value, $\zeta_c$, in order for the delocalization to occur, but it rather emerges as an attracting (singular) point as the original system of interacting charge-particles self-adjusts to percolation. It is this feature which leads to the dynamical rule of advancing the hole wave function and to a ``self-organized" formulation of the localization-delocalization transition. We reiterate that the critical strength of nonlinearity which destroys Anderson localization is expressible in terms of the percolation threshold as $\zeta_c = -1/\ln p_c$. This last result suggests that behavior be {\it non-perturbative} in vicinity of the critical point. Theoretically, this observation is very important as it elucidates the nature of thresholded phenomena including SOC. It is also manifest in the pseudochaotic character of the lattice activities, implying that the correlations persist despite that the microscopic dynamics are inherently random \cite{PRE09,Report,EPLI,PD2004}.

\subparagraph{Asymptotic spreading of the hole wave function} Let us now obtain second moments for the threshold spreading of the wave-field. This task is essentially simplified if one visualizes the transport of the hole wave function as a random walk over a system of dephased oscillators. For $p\rightarrow p_c$, this system is self-similar, i.e., fractal. It is this fractal distribution of dephased oscillators which, according to Bak, Tang, and Wiesenfeld \cite{BTW}, conducts ``{\it the noise signal$\dots$ through infinite distances}" just above the marginally stable state. We consider this distribution as a percolation cluster on a Cayley tree. The fractal geometry of this cluster is fully characterized by the Hausdorff dimension $d_f = 4$ and the index of anomalous diffusion, $\theta = 4$ (e.g., Refs. \cite{Naka,Havlin}). Note that the spectral fractal dimension is exactly 4/3 in this limit, consistenly with the original AO result \cite{AO}. From Eq.~(\ref{RW}) one obtains
\begin{equation}
\langle (\Delta n) ^2 (t) \rangle \propto t^{1/3}, \ \ \ t\rightarrow+\infty.
\label{MS++} 
\end{equation}
This behavior is asymptotic in the thermodynamic limit. Remark that the Hausdorff dimension being equal to 4 matches with the implication of Eqs.~(\ref{4ap}) and~(\ref{5ap}) where the coefficients $V_{k, m_1, m_2, m_3}$ are supposed to run over 4-dimensional subsets of the ambient mapping space. Indeed it is the overlap integral of four Anderson eigenmodes, Eq.~(\ref{5ap}), that decides on dimensionality of subsets of phase space where the transport processes concentrate. When the nearest-neighbor rule is applied, this overlap structure is singled out for the dynamics. Under the condition that the structure is critical, i.e., ``at the edge" of permitting a path to infinity, the support for the transport is reduced to a percolation cluster on a Bethe lattice $-$ characterized, along with the above value of the Hausdorff dimension, by the very specific connectivity exponent, $\theta = 4$. The end result is $2 / (2+\theta)  = 1/3$. 
  
\subparagraph{Summary} We have shown that the Anderson localization in disordered media can be lost in the presence of a weak nonlinearity and that the phenomenon is critical (thresholded). That means that there is a critical strength of nonlinearity above which the wave field turns to unlimited spreading. Below that limit, the field is localized similarly to the linear case. We have discussed this localization-delocalization transition as a percolation transition on the separatrix system of discrete nonlinear Schr\"odinger equation with disorder. This problem is solved exactly on a Bethe lattice. A threshold for delocalization is found to be $\zeta_c = 1/\ln 2 \approx 1.4427$. For the $\zeta$ values smaller than this, the localization persists, despite that the problem is nonlinear. Support for this type of behavior can be found in the results of Ref. \cite{WangF}. More so, a ``self-organized" formulation of the localization-delocalization transition has been obtained on the basis of DANSE~(\ref{DANSE}) by defining the nonlinearity parameter on-the-fly, thus offering a fertile playground to describe the phenomena of SOC. In vicinity of the delocalization point the spreading of the wave field is subdiffusive, with second moments that grow with time as a powerlaw $\propto t^{1/3}$ for $t\rightarrow+\infty$. This regime bears signatures enabling to associate it with the onset of ``weak" transport \cite{NF95} of Alfv\'en eigenmodes (AEs) in the vicinity of marginal stability of magnetic confinement systems. In this respect, we note that the characteristic aspects of sandpile physics involving SOC have, in the AE transport case, been discussed in Refs. \cite{Dendy,C&Z}.

\section{The two faces of nonlinearity: Instability of SOC}

Often when SOC systems are said to be ``nonlinear" one refers to the operation of a feedback mechanism \cite{Sornette,PT} ensuring that the control parameters need not be fine tuned explicitly to obtain criticality. It is this feedback which stabilizes the system at the state of marginal stability, or the SOC state, as opposed to traditional critical phenomena, which do require a tuning. The comprehension of nonlinear feedback mechanism leads to another face of self-organization, the existence of a bursting instability of SOC \cite{EPL,NJP}, which occurs in the parameter range of excessive external forcing, and for which we suggest the name ``fishbone-like" instability, by analogy with some bursting (internal-kink) instabilities in magnetic confinement systems \cite{Chen}. The implication is that nonlinearity can play either stabilizing as in the ideal SOC regime or destabilizing as in the regime of overdriving, role depending on the strength of interaction with the exterior. This section is aimed to discuss this topic in more detail with the aid of the DPRW model. 

\subparagraph{Instability cycle} As the probability of site occupancy $p$ approaches the percolation threshold $p_c$, the pair connectedness length diverges as $\xi \propto |p-p_c|^{-\nu}$. For $p > p_c$, the longest relaxation time in the system is $T_{\xi} \propto (p-p_c)^{-z\nu}$ and the dynamic susceptibility behaves as $\chi \propto (p-p_c)^{-z\nu\gamma}$. In the DPRW model $p$ as a function of time is determined dynamically by the competing charge deposition and loss processes. That is, $dp / dt = Z_{+} - Z_{-}$, where $Z_{+}$ is the net deposition rate of the free charges on the capacitor's left plate, and $Z_{-}$ is the particle loss rate. The net deposition rate, or the driving rate, is the control parameter of the model: It takes a given value. The particle loss rate is obtained as electric current in the ground circuit, i.e., $Z_{-} = I \theta (p-p_{\min})$, where $p_{\min}$ is the lower limit of variation of $p$. Note that $Z_{-}$ is due to the free particles leaving the system through the earthed plate. The Heaviside $\theta$ function indicates that the lattice can release charges only if/when $p\geq p_{\min}$. We expect that $p_{\min}$ lies close to, although somewhat lower than, the percolation threshold $p_c$. This is because the conducting cluster can still loose its charge content to the ground circuit even in the absence of a connecting path to the charged plate. The dynamics of $I$ can be estimated from $dI / dt = \pm I / T_{\xi}$, where the upper sign corresponds to the relaxation process in the lattice. Putting all the various pieces together, we write
\begin{equation}
dp / dt = Z_{+} - I \theta (p-p_{\min}),\label{4} 
\end{equation}
\begin{equation}
dI / dt = W I |p-p_c|^{z\nu} {\rm sign} (p-p_c),\label{5} 
\end{equation}
where ${\rm sign} (p-p_c) = + 1$ ($-1$) for $p>p_c$ ($p < p_c$) is the sign-function, and $W$ is a numerical coefficient. Equations~(\ref{4}) and~(\ref{5}) define a simple system of equations for two cross-talking variables, the lattice occupancy per site, and the particle loss current. These model equations are perhaps the simplest nonlinear equations describing the generic fueling-storage-release cycle in driven, dissipative, thresholded dynamical systems. An examination of these equations shows that the dynamics are periodic (auto-oscillatory), with the peak value of electric current $I_{\max} \simeq W (p_{\max} - p_c)^{z\nu + 1}$. Here, $p_{\max}$ ($p_{\max} > p_c$) is the upper limit of the $p$ variation. Note that $I_{\max} \rightarrow 0$ for $p_{\max} \rightarrow p_c$ as expected. The auto-oscillatory motions signify that the pure SOC state is destabilized and that the systems phase trajectories enter the supercritical parameter range. When $p_{\max} \rightarrow 1$, the periodic dynamics acquire a sharp, bursting character. The bursts half-duration (or half-width) equals $\Delta \simeq (1/W)(p_{\max} - p_c)^{-z\nu}$. Eliminating the distance to the critical state one obtains the scaling relation $I_{\max} \propto \Delta ^{-b}$, where $b = (z\nu + 1) / z\nu$. The period between the bursts is found to be $\Theta_b \simeq (p_c - p_{\min}) / Z_+$. A pure SOC state with no superimposed periodic bursts arises when $\Theta_b \rightarrow \infty$. This implies that $Z_+ \rightarrow 0$ for $p\rightarrow p_c$. Thus, criticality requires the vanishing of $Z_+$, in agreement with the result of Ref. \cite{MF}. The critical state is stable when $\Theta_b \gg T_{\xi}$. We have 
\begin{equation}
(p_c - p_{\min}) / Z_+ \gg T_{\xi} \propto |p-p_c|^{-z\nu}.\label{Stab} 
\end{equation}
This is satisfied when $Z_+ \rightarrow 0$ faster than
\begin{equation}
Z_{+\max} \propto |p-p_c|^{z\nu}.\label{Zet} 
\end{equation}
This limit exceeded, the system turns to auto-oscillate around the percolation point with a period dictated by the net deposition rate of the polarization charges. The physics origin of this auto-oscillatory motion lies in the fact that the changing amount of the free particles provides a feedback on the lattice occupancy parameter. It is due to this feedback relation that the DPRW system operates as a self-adjusting, intrinsically {\it nonlinear} dynamical system. Whether or not this feedback will excite the instability depends on how the characteristic driving time compares to the characteristic relaxation time. Indeed, focus on the stability condition in Eq.~(\ref{Stab}). The system being stable at the percolation point requires that the relaxation time due to the random walks $T_{\xi}$ be short compared to the characteristic driving time, $1/Z_+$. In this parameter range, any occasional charge density perturbations will dissipate via the random walks before input conducting sites are again introduced. When the percolation point is approached, because the time scale $T_{\xi} \propto |p-p_c|^{-z\nu}$ diverges, it is essential that the system be driven infinitesimally slowly to remain at a pure SOC state. Instability occurs when the relaxation processes operate on a longer time scale than the driving processes. In this regime, the system accumulates the polarization charges,\footnote{Remember that, in the proposed model, the polarization charges act as conducting states for the motion of current-carrying particles.} whereas to remain at criticality it would get rid of them. The accumulation of the polarization charges has direct effect on the conductivity between the plates, which steps up with the lattice overshooting the percolation threshold. When $p_{\max} \rightarrow 1$, the system can be thought of as facing the typical conditions of electrostatic discharge in the regime of short-circuit. It should be emphasized that the feedback mechanism does a two-fold job: (i) it stabilizes the system at the state of critical percolation in a regime when the driving rate is infinitesimal; and (ii) it excites a cross-talk between the conductivity and the lattice occupancy parameters when the driving rate is faster than the relaxation rate. 

In the parameter range in which the strength of the driving vanishes, the multi-scale geometry of the critical percolation is dominant in providing the major transport characteristics for the DPRW lattice. The situation changes drastically when the strength of the driving increases above some level. With the systems departure away from the percolation point, the multi-scale features will soon be lost substituted by the bulk-average nonlinearities. The fact that Eqs.~(\ref{4}) and~(\ref{5}) above are formulated in terms of the system-average parameters, $p$ and $I$, merely reflects that the system is allowed to appreciably depart from the state of marginal stability, or the SOC state (that means that $p_{\max}$ can be rather closer to 1 than to $p_c$), and that the effect of overdriving readily calls for the global features to come into play. It is noted that, in general, the multi-scale properties due to SOC can coexist along with the global or coherent features; one example of this is substorm behavior of the dynamic magnetosphere \cite{NJP}.    

The end result of the discussion above is that the strength of the driving plays a crucial role in dictating both linear and nonlinear behaviors in the DPRW model. To obtain a pure SOC state the driving rate should go to zero fast enough as the critical point is approached. The main effect overdriving has on the DPRW dynamics is to excite unstable modes associated with periodic bursts in the particle loss current. Accordingly, the system auto-oscillates between a subcritical ($p_{\min} < p_c$) and a supercritical ($p_{\max} > p_c$) states in response to external forcing. The transition to auto-oscillatory dynamics signifies the increased role of global and nonlinear behaviors in the strongly driven DPRW system as compared to a pure SOC system. The borderline between the two regimes corresponds to $T_{\xi} \simeq (p_c - p_{\min}) / Z_+$. The stability condition $T_{\xi} \ll (p_c - p_{\min}) / Z_+$ has serious implications for the achievable SOC regimes. It poses one important restriction on the net deposition rate of the free particles against the longest relaxation time on the incipient percolation cluster.

\subparagraph{``Fishbone"-like instability} To help judge the result obtained, let the critical exponents take their mean-field values: $z=2$, $\nu = 1/2$. In this limit Eqs.~(\ref{4}) and~(\ref{5}) above reproduce, up to change of variables, Eqs.~({13}) and~({14}) of Ref. \cite{Chen}. The latter set of equations appear in a basic theory of Alfv\`en instabilities as a simplified model for the coupled kink-mode and trapped-particle system in a magnetically confined toroidal plasma where beams of energetic particles are injected at high power. The mode dubbed ``fishbone" is characterized by large-amplitude, periodic bursts of magnetohydrodynamic (MHD) fluctuations, which are found to correlate with significant losses of energetic beam ions \cite{Chen}. By comparing our Eqs.~(\ref{4}) and~(\ref{5}) with Ref. \cite{Chen} one can see that the lattice occupancy per site $p$ corresponds to the effective resonant beam-particle normalized pressure within the $q=1$ surface (here, $q$ is the familiar safety factor used in tokamak research); $p_c$ corresponds to the mode excitation threshold; and the particle loss current $I$ corresponds to the amplitude of fishbone. This direct correspondence between the two models suggests consider the instability in Eqs.~(\ref{4}) and~(\ref{5}) as analog ``fishbone" instability for SOC dynamics. 

This correspondence is not really surprising. Mathematically, it stems from the resonant character of the fishbone excitation, implying that the energetic particle scattering process is directly proportional to the amplitude of fishbone \cite{C&Z,Chen}. This resonant property dictates a specific nonlinear twist to the fishbone cycle, differentiating it from other bursting instabilities in magnetically confined plasmas. It is this ``resonant" twist observed in the DPRW model system that identifies the analog ``fishbone" mode for SOC. We note in passing that the existence of an instability on the top of SOC dynamics conforms with the results of Ref. \cite{Sanchez} in which the traditional (sandpile) SOC model has been modified by adding diffusivity, giving rise to periodic relaxation-type events as a function of the system drive, while a pure SOC state requires a vanishing drive.\footnote{Beside this share, the mode referred to in Ref. \cite{Sanchez} is edge localized, non-resonant, fluid mode, in agreement with the diffusive nature of the added flux, but at contrast with the resonant mechanism of the fishbone excitation.} 

\subparagraph{The threshold character of fishbone excitation} The fishbone belongs to a specific class of instabilities, the energetic particle modes (EPM), which appear in a magnetic confinement system when the energetic particle pressure is comparable with the pressure of the thermal plasma \cite{C&Z,EPM}. The EPM constitutes a separate branch with a distinctive dispersion relation and its frequency and the growth rate crucially depend on the parameters of the energetic particle orbital motion. When the drive is strong enough, the EPM may be unstable despite having a frequency in the Alfv\`en continuum where normal modes of the background plasma are typically strongly damped (the associated damping rate is proportional to the gradient of the phase velocity) \cite{Heid}. Instabilities in the Alfv\`en continuum are often observed during intense neutral-beam injection \cite{Chen,Coppi}, but they can also be excited by the energetic electrons generated experimentally by different means: electron cyclotron resonance heating (ECRH) as on DIII-D tokamak \cite{Wang} and lower hybrid (LH) power injection as in Frascati Tokamak Upgrade (FTU) experiments \cite{NF2007}. The phenomenon is thresholded in that it requires a critical level of the absorbed power. The existence of the critical power, which we associate with the critical ``driving" rate, $Z_{+\max}$, is well established experimentally (Fig.~8).

\begin{figure}
\includegraphics[width=1.00\textwidth]{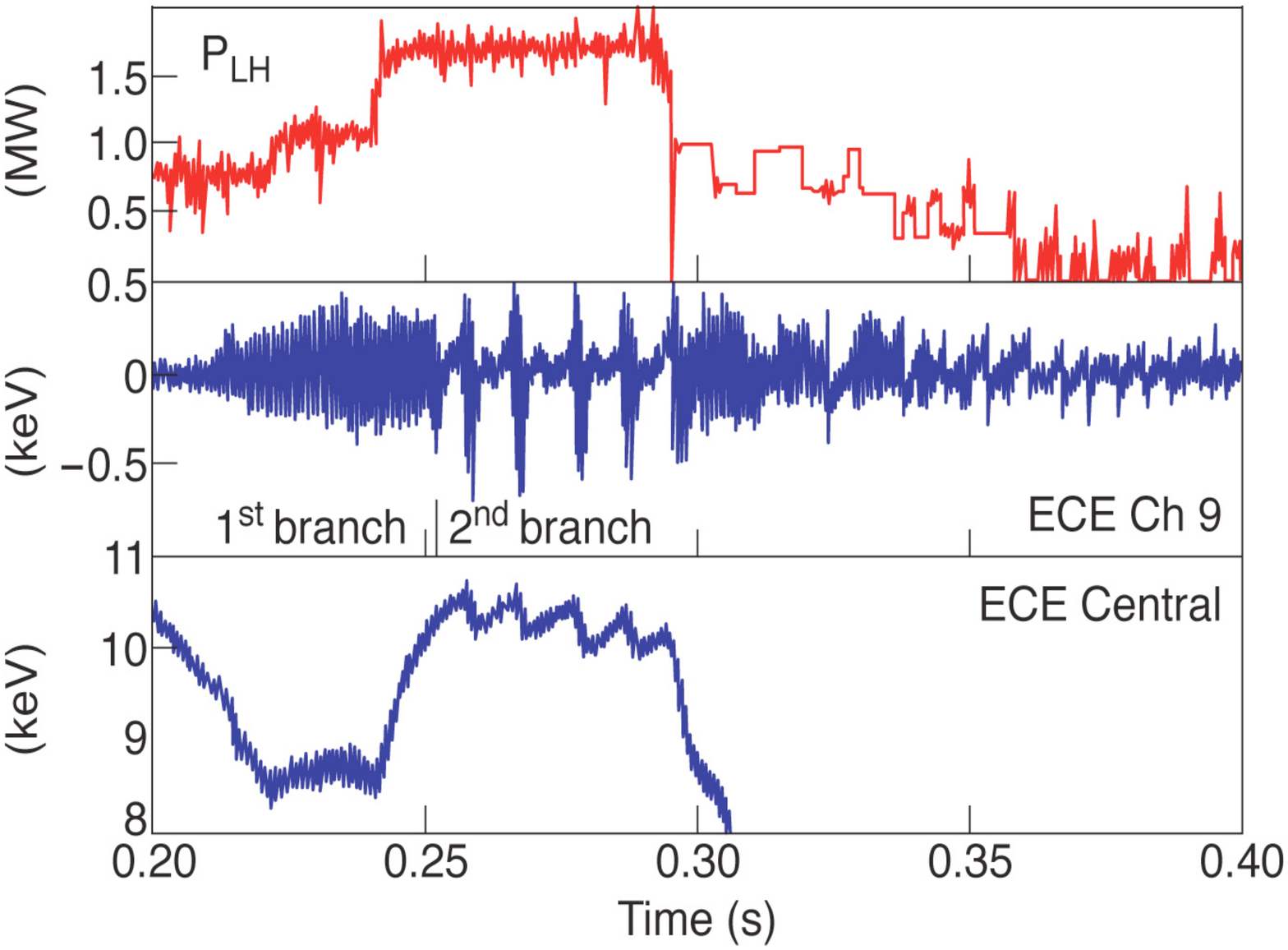}
\caption{\label{} Plots of LH coupled power, fast electron temperature fluctuations, and central radiation temperature in FTU shot $\#$ 20865. During high power LH injection, an evident transition in the electron fishbone signature takes place from almost steady state nonlinear oscillations (fixed point; marked as $1^{\rm st}$ branch) to regular bursting behavior (limit cycle; marked as $2^{\rm nd}$ branch). The transition is at $P_{\rm LH} \approx 1.69$ MW. It is noticed that the bursting behavior phase closely resembles that of well-known ion fishbones \cite{Chen,Coppi} and ECRH driven electron fishbones on DIII-D \cite{Wang}. Adapted from Ref. \cite{NF2007}.}
\end{figure}

The EPM dispersion relation \cite{NF95,C&Z,Chen,EPM} when account is taken for the well-known ``fluid" (this includes the background MHD and the energetic particle adiabatic and convective responses) and ``kinetic" contributions due to the energetic-particle ``compressions" can be obtained via asymptotic matching procedure, leading, upon the fast and slow time scales are separated, to the frequency-dependent complex nonlinear parabolic equation, Eq.~(15) of Ref.~\cite{NF95}. A remarkable feature of this equation is that the nonlinearity due to the wave field is twisted with the free energy source term (here thought as the ``driving" term). The main effect this twist has on the dynamics is that the EPMs are released in radially amplifying ``avalanches" \cite{C&Z} (to visualize, think to a large mass of mud or snow moving rapidly downhill). This avalanching behavior which we associate with behavior of a strongly overly driven system in the presence of intense energetic particle population should be distinguished from the above ``chain reactions of hopping motions," which are the avalanches in the DPRW SOC system at criticality. In the {local limit}, characterized by a Gaussian free energy source profile, Eq.~(15) of Ref.~\cite{NF95} is reduced to a complex NLSE, which is different from DANSE~(\ref{DANSE}) in that it is dominated by the nonlinear properties, rather than by a competition with randomness. 

\subparagraph{Fractional nonlinear Schr\"odinger equation} If one wants to go beyond the local limit, one may use a stretched Gaussian free energy profile instead \cite{PPCF}. In the latter case, the resulting equation is found to be a variant of fractional nonlinear Schr\"odinger equation, or FNLSE (presented here without derivation, see Ref. \cite{UFN}) 
\begin{equation}
i\frac{\partial \Psi (t,x)}{\partial t} - {\mathcal{D}}_q \nabla_{-x}^{q}\nabla_x^{q}\Psi (t,x) + \zeta |\Psi (t,x)|^2 \Psi (t,x) = 0,\label{FNLS} 
\end{equation}
where 
\begin{equation}
\nabla_x^{q}\Psi (t,x) = \left[1/\Gamma (1-q)\right] x^{-q}\star\Psi (t,x) \equiv \left[1/\Gamma (1-q)\right] \int_{-\infty}^{+\infty} dy |x-y|^{-q} \Psi (t,y)\label{Convol} 
\end{equation}
is the so-called Riesz/Weyl fractional derivative \cite{Klafter,Podlubny}, which is mathematically different from the Riemann-Liouville derivative in Eq.~(\ref{23}) in that the integration is performed through infinite limits; the operator $\nabla_{-x}^{q}\nabla_x^{q}$ is a generalization of the Laplacian; $x$ denotes respective spatial coordinate; $q$ is a fractional exponent ($0\leq q<1$) which corresponds with the exponent of the stretched Gaussian free energy source profile \cite{PPCF}; and ${\mathcal{D}}_q$ is a normalization constant which carries the dimension $\left[{\mathcal{D}}_q\right] = {\rm cm}^{2q}\cdot{\rm s}^{-1}$ (we assume that $\hbar = 1$). For attractive nonlinearity (here corresponding to $\zeta > 0$), FNLSE~(\ref{FNLS}) describes phenomena of self-delocalization of fractons (vibrational excitations of fractal networks) as well as associate phenomena of nonlocal solitary waves \cite{UFN}. Setting $\Psi (t,x) = \psi (x) \exp (-i\omega t)$, from FNLSE~(\ref{FNLS}) one arrives at the fractional envelope equation
\begin{equation}
- {\mathcal{D}}_q \nabla_{-x}^{q}\nabla_x^{q}\psi (x) + \omega \psi (x)  + \zeta |\psi (x)|^2 \psi (x) = 0,\label{Envel} 
\end{equation}
which does not contain time differentiation. Remark that FNLSE~(\ref{FNLS}) is built on fractional derivatives in {\it space}, while time differentiation is conventional ({integer}), likewise to DANSE~(\ref{DANSE}). This distinction between the fractional derivatives in space and time reflects the very special role \cite{UFN,Comb} time plays in the dynamical equations originating from quantum mechanics such as NLSE, by contrast with the kinetic equations for transport and relaxation processes discussed above \cite{Klafter,Ctrw1,Ctrw2}.  

\subparagraph{Mixed SOC-coherent behavior} The idea of ``fishbone" instability in self-organized critical dynamics is very appealing as it addresses a type of behavior in which the multi-scale features due to SOC can coexist along with the global or coherent features. One example of this coexistence can be found in solar wind$-$magnetosphere interaction. Indeed it has been discussed by a few authors \cite{Chang,Klimas,Uritsky,Kozelov} that the coupled solar wind$-$magnetosphere$-$ionosphere system operates as an avalanching system and that there is a significant SOC component in the dynamics of magnetospheric storms and substorms, along with a coherent component \cite{Chang,Sharma} that evolves predictably through a sequence of clearly recognizable phases \cite{Baker}. Here, we advocate a way of thinking \cite{UFN,JASTP} in which the magnetospheric SOC component is associated with the properties of self-organization of electric currents and magnetic field fluctuations in the plasma sheet of the Earth's magnetotail (i.e., the ``Sakura" model); whereas the coherent component is attributed to global instability of the cross-tail SOC current system and the phenomena of tail current disruption. The implication is that the dynamic magnetosphere survives through a {\it mixed} SOC-coherent behavior. In this spirit, we expect the input power due to magnetic reconnection at the Earth's dayside magnetopause to self-consistently control the departure from the state of marginal stability, or the SOC state, with stronger departures favoring the coherent features. By analogy with fishbone instability in magnetic confinement systems we suggest that behavior is thresholded in that there is a critical input power (critical reconnection rate at the dayside magnetopause), destabilizing the SOC component in the magnetotail. At this point, a portion of the cross-tail electric current will be redirected to ionosphere $-$ thought as analog ``ground circuit" (Fig.~9) $-$ leading to a decrease in the tail current intensity (tail current disruption), and to a magnetospheric disturbance, or a substorm. A model for the substorm cycle is obtained from the above coupled system of equations, Eq.~(\ref{4}) and~(\ref{5}), where one identifies the driving rate, $Z_{+}$, with the magnetic reconnection rate at the Earth's dayside magnetopause; the lattice occupancy parameter, $p$, with the average normalized energy density of magnetic fluctuation field in the magnetotail current sheet; and the particle loss current, $I$, with electric current in the ionosphere. We should stress that we consider a substorm as instability in the cross-tail SOC current system, which occurs on the top of self-organization to a critical state in the magnetotail, thus posing a coherent component over the dynamics. 

\begin{figure}
\includegraphics[width=1.10\textwidth]{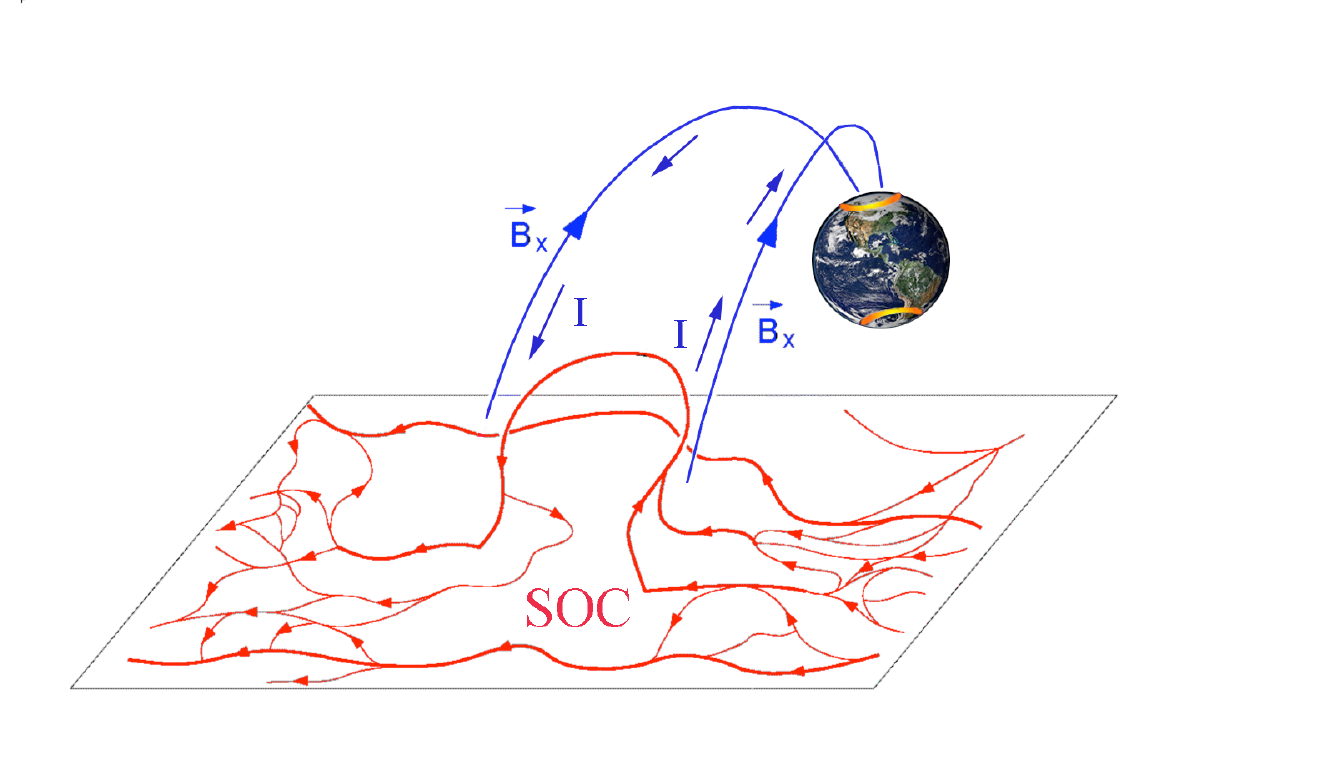}
\caption{\label{} Substorm in the Earth's magnetosphere. Red: The complex electric current system in the magnetotail current sheet. Blue: The magnetotail lobe field, ${\bf B}_x$. A current filament popping up in the plasma sheet interacts with the Harris-distributed magnetic field in the lobes of the magnetotail. The forces are such as to make the filament spontaneously change its orientation, favoring redirection of electric current to the ionosphere.}
\end{figure}

\section{Phase transitions in SOC systems }  

It was argued that the phenomena of magnetospheric substorm bear signatures enabling to associate them with a second-order phase transition in the coupled perturbation electric current and magnetic fluctuation system \cite{JGR,JASTP}. Indeed, when a current filament pops up in the plasma sheet, it interacts with the Harris-distributed\footnote{According to Harris \cite{Harris}, the dependence of the lobe field is given by hyperbolic tangent of the distance to the neutral plane.} magnetic field in the lobes of the magnetotail (Fig.~9). The forces are such as to make the filament spontaneously change its orientation, and in fact a local minimum in the free energy profile occurs, which favors current disruption to reinstall stability. 

Let us address the phenomena of magnetospheric substorm from a more fundamental perspective, namely, as part of the general problem of phase transitions in SOC systems. The main idea here is that some systems may spontaneously turn into a coherent state before they become SOC, since their evolution by itself drives these system to a competition between the SOC and coherent properties as a consequence of some nonlinear twist between associate order parameters. Other than substorms, this general approach may include phenomena like the L-H transition in magnetic confinement devices \cite{Jeffrey}, and the tokamaks as particular case, where the L-phase is associated with SOC \cite{Carreras}, and the H-phase is associated with spontaneously occurring coherent state.  

\subparagraph{Subordination to SOC} Let us consider a spatially extended system with some order parameter $\Upsilon$, where the processes of self-organization develop a singularity at some value $\Upsilon_c$ for $t\rightarrow+\infty$. We assume that this singularity does not explicitly appear in the dynamics, implying that the system is, in this limit, critical and self-organized. The phenomena we are looking at appear when the system possesses a {\it competing} order parameter, which we shall denote by $\psi$, and for which $\Upsilon$ acts as input control parameter. The implication is that the dynamics of $\psi$ is subordinated to the dynamics of $\Upsilon$ via some intrinsic coupling mechanism. For simplicity, we shall assume that the order parameter $\psi$ corresponds to a coherent behavior, which we envisage as competing with the emerging multi-scale features due to SOC. Thus, while the system is developing its singular (SOC) points, it may find it thermodynamically profitable to spontaneously turn into the competing, coherent phase. As we shall see, this idea leads to a fractional extension of the Ginzburg-Landau equation \cite{UFN,PLA}, in which the conventional Laplacian is replaced by  fractional Riesz/Weyl derivatives.     

\subparagraph{Generalized free energy expansion} For the sake of mathematical convenience, we shall assume that the system approaches its SOC point so closely, that the geometry is, to a good approximation, self-similar (fractal). In this regime, the distribution of the competing order parameter $\psi$ will be characterized by the diverging correlation length, owing to its coupling with $\Upsilon$, so that both $\psi$ and $\Upsilon$ distributions are heavy-tailed. In a sense, the fractal distribution of $\Upsilon$ acts a fractal support for $\psi$. The fact that the $\psi$ variation involves correlations on many spatial scales must have implications for the generalized form of the free energy expansion near the phase transition point, and in particular for the gradient term, where the usual assumptions of locality, permitting to write this term as a simple $|\nabla_x\psi (x)|^2$, should be relaxed. Then a consistent generalization accounting for the integral effect of the correlations is obtained in terms of a Fourier convolution, $\nabla_x^{q}\psi (x) = \left[1/\Gamma (1-q)\right] x^{-q}\star\psi (x)$, or the Riesz/Weyl fractional derivative \cite{Klafter,Podlubny}, where $q$ ($0 < q \leq 1$) characterizes the strength of spatial correlation and the local limit is reinstalled for $q\rightarrow 1$. It is this convolution which we expect to replace $\nabla_x\psi (x)$ when the gradient term is considered. Indeed, the following generalized free energy expansion in vicinity of the transition point holds \cite{UFN,PLA} 
\begin{equation}
F=F_0 + \int_{-\infty}^{+\infty} dx \left[{\mathcal{A}}_q |\nabla_x^{q}\psi (x)|^2 + a_q |\psi (x)|^2 + \frac{1}{2}b_q |\psi (x)|^4\right],\label{FEE} 
\end{equation}
where we have introduced three phenomenological expansion parameters ${\mathcal{A}}_q$, $a_q$, and $b_q$, which may depend on the exponent $q$ in general. 

\subparagraph{Fractional Ginzburg-Landau equation} Varying the functional in Eq.~(\ref{FEE}) over the complex conjugate $\psi^* (x)$ and considering $\psi (x)$ and $\psi^* (x)$ as independent order parameters, one obtains
\begin{equation}
\delta F=\int_{-\infty}^{+\infty} dx \left[{\mathcal{A}}_q \nabla_x^{q}\psi (x)\nabla_x^{q}\delta\psi^* (x) + a_q \psi (x)\delta\psi^* (x) + b_q |\psi (x)|^2\psi (x)\delta\psi^* (x)\right].\label{Var} 
\end{equation}
With use of the integration-by-parts formula \cite{Podlubny}
\begin{equation}
\int_{-\infty}^{+\infty} dx \varphi_1 (x) \nabla_x^{q}\varphi_2 (x) = \int_{-\infty}^{+\infty} dx \varphi_2 (x) \nabla_{-x}^{q}\varphi_1 (x)\label{Byp} 
\end{equation}
equation~(\ref{Var}) implies that  
\begin{equation}
\delta F=\int_{-\infty}^{+\infty} dx \left[{\mathcal{A}}_q \nabla_{-x}^{q}\nabla_x^{q}\psi (x) + a_q \psi (x) + b_q |\psi (x)|^2\psi (x)\right]\delta\psi^* (x),\label{Var+} 
\end{equation}
yielding, in view of the extremum $\delta F = 0$,
\begin{equation}
{\mathcal{A}}_q \nabla_{-x}^{q}\nabla_x^{q}\psi (x) + a_q \psi (x) + b_q |\psi (x)|^2\psi (x) = 0.\label{FGL} 
\end{equation}
Note that varying the integral in Eq.~(\ref{FEE}) over $\psi (x)$ leads to the conjugate equation
\begin{equation}
{\mathcal{A}}_q \nabla_{-x}^{q}\nabla_x^{q}\psi^* (x) + a_q \psi^* (x) + b_q |\psi (x)|^2\psi^* (x) = 0,\label{FGL+} 
\end{equation}
which is physically identical to Eq.~(\ref{FGL}). Equation~(\ref{FGL}) has the mathematical structure of the well-known Ginzburg-Landau equation \cite{PartII}, in which the conventional Laplacian, $\nabla_x^2$, is generalized to $\nabla_{-x}^{q}\nabla_x^{q}$. We consider Eq.~(\ref{FGL}) as fractional Ginzburg-Landau equation, or FGLE. One sees that FGLE appears as a natural tool in describing phase transitions in SOC systems, in much the same way as the conventional Ginzburg-Landau equation describes type II phase transitions in simpler systems. Remark that FGLE~(\ref{FGL}) is different from the fractional envelope equation, Eq.~(\ref{Envel}), in that it contains the opposite sign in front of $\nabla_{-x}^{q}\nabla_x^{q}$. 

\subparagraph{The $q$ exponent} Likewise to traditional type II phase transitions one may argue that $a_q$ changes sign at the critical point and that it linearly depends on variation of the input control parameter \cite{PartII,PartI}. Then the subordination condition will imply that $a_q = \alpha_q (\Upsilon - \Upsilon_c)$ for $\Upsilon \rightarrow \Upsilon_c$, with $\alpha_q$ a constant which does not depend on $\Upsilon$. Given that the system is driven so slowly that it develops a singular point as a result of self-organization, one predicts, with the use of FGLE~(\ref{FGL}), that the distribution of the order parameter $\psi$ will be self-similar to comply with the scaling $|\psi (x)|^2 \propto x^{-2q}$, from which the Hausdorff dimension $d_f = d - 2q$ can be deduced. Focusing on the $d_f$ value, because the order parameter $\psi$ is intrinsically coupled to $\Upsilon$, one may expect that the $\psi$ and $\Upsilon$ distributions will be essentially the same and hence, when account is taken for the percolation fractal geometry of SOC, will be both characterized by the ``hyperuniversal" relation, $d_f = d-\beta/\nu$ (see section~1) where $\beta$ and $\nu$ are percolation critical exponents. The latter expression will be consistent with FGLE~(\ref{FGL}), when $q = \beta / 2\nu$. This is the desired result. It shows that there is nontrivial fractional index of integro-differentiation in associate fractional Ginzburg-Landau equation and that behavior is nonlocal in general. Using known estimates for the parameters $\beta$ and $\nu$ \cite{Stauffer,Naka}, it is found that $q=5/96$ for $d=2$; $q\approx 0.26$ for $d=3$; and $q\approx 0.4$ for $d=4$. The mean-field value, holding for $d\geq 6$, is $q=1$. One sees that the mean-field case is local, as it should. Allowing for $q\rightarrow 1$ in FGLE~(\ref{FGL}), one verifies that the conventional (``integer") Ginzburg-Landau equation is readily reinstalled. By contrast, for $q<1$, the dynamics are governed by an interplay between nonlocality and nonlinearity, likewise to the EPM excitation case, Eq.~(\ref{FNLS}), and their correct description requires fractional extensions of respectively the Ginzburg-Landau and nonlinear Schr\"odinger equations consistently with the implication of the Riesz/Weyl fractional operator.

\section{Overall summary and final remarks}

The concept of self-organized criticality, or SOC, proves to be a complementary tool in drawing a physical picture of the processes underlying the dynamics of systems with many coupled degrees of freedom (i.e., the ``complex" systems). In this report we have demonstrated the diversity of appropriate mathematical methods of describing such processes, including fractional equations of the diffusion, relaxation, and Ginzburg-Landau (nonlinear Schr\"odinger) type, generalizing their standard counterparts, as well as the formalism of discrete Anderson nonlinear Schr\"odinger equation (DANSE) $-$ extending far beyond the usual scaling theories. Some connections to Hamiltonian models, paving the way to first-principle models of SOC phenomena, have been also discussed. These issues make the mathematical formalism of SOC an exciting and challenging problem. 

The main emphasis in the present work has been laid on percolation, recognized as a convenient and powerful framework in describing critical phenomena in complex systems. The percolation problem finds its significance in its relation with the fundamental topology (in terms of connectedness issues) and theory of fractional manifolds \cite{UFN,PRE97}. In this respect, some elements of conformal maps of fractals, along with the percolation problem on the Riemann sphere, have been addressed. 

To deal with dynamical problems involving feedback between the various degrees of freedom such as the SOC problem we used the idea of random walks on a self-adjusting percolation cluster. This idea is advantageous, as it makes it possible to employ the random walks in place of the usual lattice automated redistribution rules, and in this fashion to design a family of SOC models, that are more friendly from the standpoint of their analytical treatment as compared to their CA relatives. In fact, the random walks on percolation systems offer suitable analytical forms for the diffusion, charge-conduction, and the dynamic susceptibility properties, and their significance in the study of SOC phenomena can hardly be exaggerated. 

More so, we proposed a simple lattice model of self-organized criticality, the DPRW model, which addresses the SOC problem as a transport problem for electric charges (free particles and holes) on a dynamical geometry of the threshold percolation. The novel concepts of this model are: (i) a theory of self-organized criticality based on the analogy with dielectric-relaxation phenomena in self-adjusting random media, and (ii) prediction of a ``resonant" instability of SOC due to the nonlinearities present. The system adjusts itself to remain at the critical point via the mechanisms of hole hopping associated with the random walk-like motion of lattice defects on a self-consistently evolving percolation cluster. 

With the random walk guide to lattice dynamics we could derive fractional analogs of the diffusion and relaxation equations, demonstrating the existence of multi-scale relaxation processes and of a broad distribution of durations of relaxation events. In particular, we have shown that the relaxation to SOC of a slightly supercritical state is described by the Mittag-Leffler relaxation function, Eq.~(\ref{ML}), and not by a simple exponential function as for standard relaxation. The ideal SOC state requires that the driving rate goes to zero faster than a certain scaling law as the percolation point is approached. The model belongs to the same universality class as the BTW sandpile, and should be distinguished from the DP-like SOC models.

Thinking of holes as ``excitations" of the marginally stable state, we considered a transport problem for the hole wave function in the context of DANSE equation with random potential on a lattice. An important feature which arises in this approach is competiotion between nonlinearity and randomness. It was argued that above a certain critical strength of nonlinearity the Anderson localization of the hole wave function was destroyed and unlimited subdiffusive spreading of the wave field along the lattice occurred. This subdiffusion is asymptotic \cite{EPLI}. We have seen that this problem of the critical spreading was intimately related with the outstanding problem of transport along separatrices in large systems \cite{ChV}. With the recognition that the transition to unlimited spreading could be described as a percolation transition on a Cayley tree, a ``self-organized" formulation of the phenomena of localization-delocalization in the presence of nonlinearity has been proposed. The results of this investigation have demonstrated the versatility of the DANSE formalism, which we believe to capture the essential key elements of self-organized critical phenomena, thus offering a general analytical framework for SOC. 

Overdriving the DPRW system near self-organized criticality was shown to have a destabilizing effect on the SOC state. The fundamental physics of this instability consists in the following. Because of rapid accumulation of the conducting sites, the system departs from the percolation point, and its geometry nonlinearly changes from fractal-like to crystalline-like. At this point, the conductivity of the system has greatly increased. As the lattice conducts more electricity, losses increase in the ground circuit. However, because the particle loss current has feedback on the lattice occupancy parameter, a cross-talk is excited between the systems average conductivity response and the distance to the critical state. We have observed that the instability cycle is qualitatively similar to the excitation of the internal kink (``fishbone") mode in tokamaks with high-power beam injection (the lattice occupancy per site $p$ corresponds to the effective resonant beam-particle normalized pressure within the $q=1$ surface; $p_c$ corresponds to the mode excitation threshold; and the particle loss current $I$ corresponds to the amplitude of fishbone). The instability is ``resonant" in that the particle loss process is directly proportional to $I$. This resonant property dictates a specific nonlinear twist to the fishbone cycle, differentiating it from other bursting instabilities in magnetically confined plasmas. 

The excitation of ``fishbone" instability in SOC systems leads to a type of behavior in which the multi-scale features due to SOC can coexist along with the global or coherent features (i.e., mixed SOC-coherent behavior). One example of this coexistence is found in the solar wind$-$magnetosphere interaction. We expect the concept of mixed SOC-coherent behavior be the plausible statistical picture for thresholded, dissipative, nonlinear dynamical systems in the parameter range of nonvanishing external forcing. In this respect, we suggest that some of the ``extreme" events, or system-scale responses, observed in complex natural and social systems \cite{Extreme} may, in fact, be the fishbone-like instabilities of SOC predicted by the present theory. 

It has been shown that some systems may spontaneously turn into a coherent state before they become SOC, since their evolution by itself drives these system to a competition between SOC and coherent properties as a consequence of some nonlinear twist between associate order parameters. We discussed a generalized free energy expansion for a system with extended spatial degrees of freedom, in which the order parameter due to SOC acts as input control parameter for the competing coherent behavior. Based on this expansion $-$ which has involved the Riesz/Weyl fractional operator in place of the standard (local) gradient $-$ a fractional generalization of the well-known Ginzburg-Landau equation has been obtained variationally \cite{PLA}. With the fractional Ginzburg-Landau equation, it should be possible to describe the much observed L-H transition in magnetic confinement devices \cite{Jeffrey}, as well as the phenomena of magnetospheric substorm and tail-current disruption.  

We believe it will be worthwhile to pursue the above considerations not only because they arise naturally in a basic theory we are considering, but also because questions of this kind might have feedback on seemingly very diverse phenomena beyond the specialized physics context of this study. We illustrate this on two examples from respectively finance and climate dynamics \cite{Climate,Extreme}. Indeed collective phenomena in finance include as partial cases the bursting of speculative bubbles, market crashes, debt contamination (now-deepening in the euro-zone), and the spreading of bankruptcy and insolvency. A simplified toy-model here might be constructed as a variant of the DPRW model discussed above, with clusters of polarization charges thought as asset market; holes thought as insolvency; and the phenomena of hole-hopping thought as debt spreading. In this context, we might predict that, when the market is de-regulated (i.e., the dynamics are random walk-like), periodic financial crises are virtually unavoidable and that the period between crashes is inversely proportional to the rate at which speculative (not absorbed by the real economy) capital flows into the market, thus ``overheating" the system.

Likewise to finance, many exciting questions such as the above arise in climate research and it would be of interest to study them from the more general perspective. The focus here is on existing periodic oscillations in Earth's global climate system, as for instance the El Ni\~{n}o/La Ni\~{n}a-Southern Oscillation, or ENSO, which is a two to seven year quasi-periodic climate pattern associated with warming ({El Ni\~{n}o}) and cooling ({La Ni\~{n}a}) of the ocean surface layer across the tropical eastern Pacific. The extremes of this oscillation $-$ which is identifiable in the climate reconstructions since thousands of years $-$ are blamed for the severe weather conditions affecting climate, habitats, and the economies in many regions of the world \cite{Mc}. In addition, ENSO involves \cite{Nino} interactions extending through different time scales with various climate phenomena, such as the seasonal cycle, interseasonal oscillations, and/or decadal oscillations \cite{Nino,NA}. 

It was argued \cite{OAI} that ENSO is a naturally occurring, oscillatory mode of the coupled ocean-atmosphere system. We suggest that this mode is strongly driven in that it is excited when atmospheric forcing exceeds a certain critical level. Then the oscillation is of ``fishbone" type, since the rules by which the mode is sustained will include features of unstable SOC dynamics \cite{NJP}. One might expect that a simplified yet relevant toy-model here will take the form of two cross-talking variables, Eqs.~(\ref{4}) and~(\ref{5}), where $p$ stands for air surface pressure and $I$ stands for ocean surface temperature and that the period of the oscillation will be inversely proportional with atmospheric forcing strength. More advanced models might refer to FNLSE~(\ref{FNLS}) with an account for cross-scale couplings between the various wave processes involved. Analysis of this general area remains to be carried out.

\begin{figure}
\includegraphics[width=1.00\textwidth]{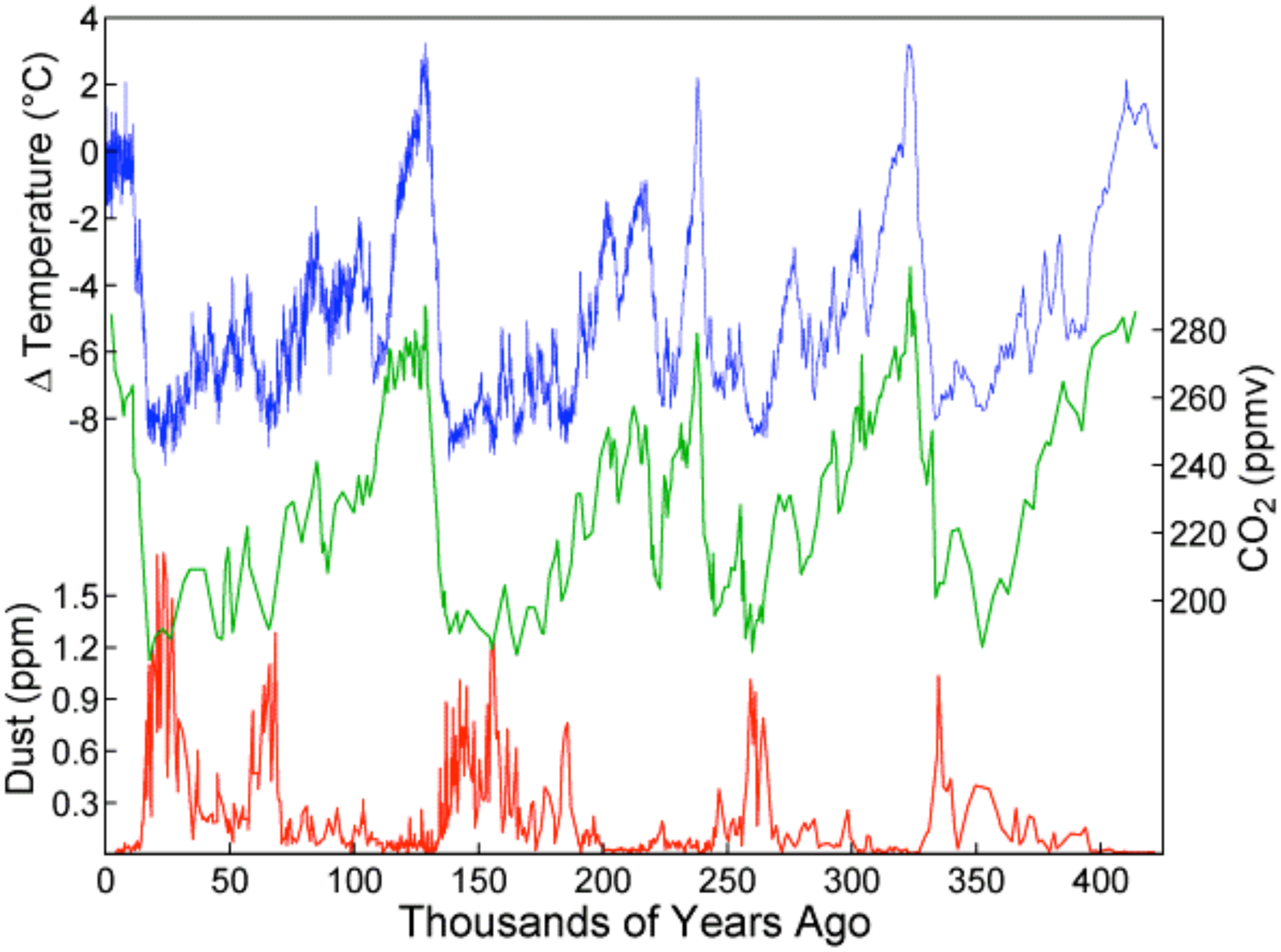}
\caption{\label{} Climate reconstruction over the past 420,000 years from the Vostok ice core, Antarctica. Blue: Temperature variation in degrees Celsius. Green: Carbon dioxide (CO$_2$) concentration in parts-per-million-by-volume, or p.p.m.v. Red: Dust concentration (dust concentrations are expressed in parts-per-million, or p.p.m., assuming that Antarctic dust has a density of $2,500$ kg$\cdot$m$^{-3}$). One sees that atmospheric concentrations of carbon dioxide correlate well with Antarctic air-temperature throughout the record, while anit-correlate with the dust concentration. Image and data credit: Ref. \cite{AntIce}.}
\end{figure}

In the same spirit, the phenomenon of fishbone might shed new light on glacial-interglacial climate changes, in particular, the $\sim$100,000-year climate cycle, which could be a globally induced unstable climate mode stemming from a cross-talk between air-temperature and dust concentration (see Fig.~10). Support for this suggestion can be found in Antarctic ice records as discussed in Ref. \cite{AntIce}. The implication is that glaciations occur naturally through functioning of Earth's climate as complex system, thus being ``inherently there" as the many degrees of freedom twist and couple with the exterior. All in all, the phenomenon of fishbone warns of repeating severe events being virtually unavoidable in driven systems.

\begin{figure}
\includegraphics[width=1.10\textwidth]{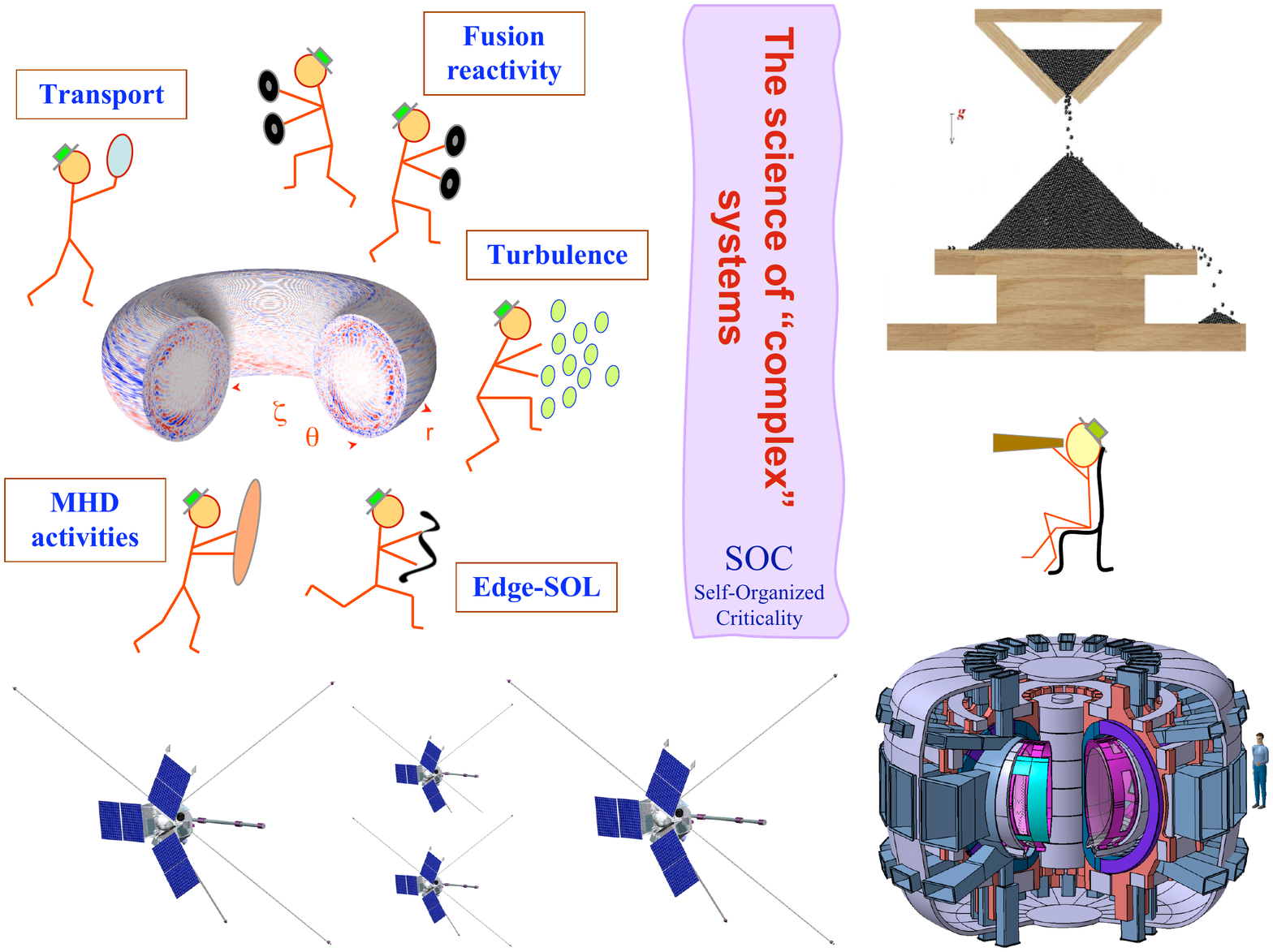}
\caption{\label{} A cartoon illustrating the many co-going plasma processes in a toroidal magnetic confinement system, viewed through the prism of complexity and self-organized criticality. A remake of Figure~1 from Ref. \cite{UFN}. In the bottom-right corner is an artist's view of the FAST tokamak. Courtesy of dr. Aldo Pizzuto. In the top-right corner is a ``sandpile" watch: Time is running for fusion. The four satellites in the bottom centre represent the ROY project \cite{ROY}.}
\end{figure}

Generally speaking, the study of self-organized criticality phenomena is currently transitioning from an emphasis on scaling and linear-response theories to an emphasis on understanding and predicting the nonlinear dynamics of systems with many coupled degrees of freedom. In many ways these tendencies are manifest in the introduction of DANSE, FNLSE, and FGLE equations discussed above. As this transition occurs, investigations $-$ as much experimental as theoretical and numerical $-$ that pinpoint the nonlinear interactions in complex systems will increase in importance. In the geo-space plasma research, with the observations becoming multi-spacecraft and/or multi-point in scope, theoretical models are likewise to confront issues of nonlocality, self-organization, and build-up of correlations in the presence of many co-existing plasma processes \cite{UFN,ROY}. Similarly to geo-space exploration, research activities in fusion plasma are now arriving at a crucial juncture that necessitates the understanding of ``complexity" in the accessible and relevant operation regimes of burning plasma. Indeed it is becoming clear that the important questions that will be receiving attention in the coming years, particularly with the development of ITER and DEMO scenarios, are addressed toward the comprehension of burning plasma state as being self-organized, thresholded, nonlinear dynamical system with many interacting degrees of freedom \cite{NF95,C&Z,PPCF,RMP}. The ITER project is a major challenge on the way to controlled fusion burn. The specialized issues of complexity, nonlinear interactions, and SOC have found their significance in the recently formulated Fusion Advanced Studies Torus, or FAST, proposal \cite{FAST}, promoting an European Union satellite for ITER. An emergent way of thinking here is to recognize FAST as having a parallel in the geo-space exploration, the ROY mission concept \cite{ROY} $-$ a project in space research for a constellation of small, probe-like satellites,\footnote{ROY means ``Swarm" in Russian} aiming to investigate the dynamic magnetosphere as complex system (see Fig.~11). We extrapolate that the cross-disciplinary effort of bringing these exciting projects to realize will open new avenues in the study of what proves to be one of the greatest theoretical challenges in the modern nonlinear physics, the paradigm of self-organized criticality.

%
%
%
%
%
\begin{table}
\caption{Abbreviations used}
\label{tab:1}       
%
%
\begin{tabular}{p{3cm}p{8.3cm}}
\hline\noalign{\smallskip}
Abbreviation & Expansion   \\
\noalign{\smallskip}\svhline\noalign{\smallskip}
AE & Alfv\'en eigenmode \\
AMPTE & Active Magnetospheric Particle Tracer Explorers (satellite) \\
AO & Alexander-Orbach (conjecture) \\
BTW & Bak-Tang-Wiesenfeld \\
CA & Cellular Automation \\
CTRW & Continuous Time Random Walks \\
ECRH & Electron Cyclotron Resonance Heating \\
ENSO & El Ni\~{n}o/La Ni\~{n}a-Southern Oscillation \\
EPM & Energetic Particle Mode \\
DANSE & Discrete Anderson Nonlinear Schr\"odinger equation \\
DEMO & DEMOnstration Power Plant \\
DIII-D & DIII-D tokamak \\
DP & Directed Percolation \\
DPRW & Dynamic Polarization Random Walk \\
ITER & International Thermonuclear Experimental Reactor \\
FAST & Fusion Advanced Studies Torus \\
FDE & Fractional Diffusion equation \\
FGLE & Fractional Ginzburg-Landau equation \\
FNLSE & Fractional Nonlinear Schr\"odinger equation \\
FTU & Frascati Tokamak Upgrade \\
KAM & Kolmogorov-Arnold-Moser \\
KWW & Kohlrausch-Williams-Watts (relaxation function) \\
LH & lower hybrid (oscillation) \\
L-H & low-high (transition) \\
MHD & Magnetohydrodynamic \\
MW & Megawatt \\
NLSE & Nonlinear Schr\"odinger equation \\
ROY & (``Swarm") IKI-led multi-satellite geo-space mission concept \\
SOC & Self-Organized Criticality \\
\noalign{\smallskip}\hline\noalign{\smallskip}
\end{tabular}
\end{table}
\begin{acknowledgement}
Valuable discussions with A. Iomin, J. Juul Rasmussen, K. Rypdal, L. M. Zelenyi, G. Zimbardo, and F. Zonca are gratefully acknowledged. This work was supported by the Euratom Communities under the contract of Association between EURATOM/ENEA.
\end{acknowledgement}

\end{document}